\documentclass[prb,twocolumn,superscriptaddress,floatfix,noshowpacs,10pt,longbibliography]{revtex4-2}%

\usepackage{graphicx,bm,times}
\usepackage{amsmath}
\usepackage{amsfonts}
\usepackage{amssymb}

\usepackage{bm}
\usepackage{color}
\usepackage{times}
\usepackage{lipsum}

\renewcommand{\vec}[1]{\mathbf{#1}}

\DeclareUnicodeCharacter{2212}{-}

\usepackage[%
colorlinks=true,
urlcolor=blue,
linkcolor=blue,
citecolor=blue
]{hyperref}

\begin{document}

\title{Resurgence of superconductivity and the role of $d_{xy}$ hole band in FeSe$_{1-x}$Te$_x$ }

\author{Archie B. Morfoot}
\email[corresponding author: ]{archie.morfoot@physics.ox.ac.uk}
\affiliation{Clarendon Laboratory, Department of Physics,
University of Oxford, Parks Road, Oxford OX1 3PU, UK}
\affiliation{Diamond Light Source,  Harwell Science and Innovation Campus, Didcot, OX11 0DE, UK}

\author{Timur K. Kim}
\affiliation{Diamond Light Source,  Harwell Science and Innovation Campus, Didcot, OX11 0DE, UK}

\author{Matthew D. Watson}
\affiliation{Diamond Light Source, Harwell Science and Innovation Campus, Didcot, OX11 0DE, UK}

\author{Amir A. Haghighirad}
\affiliation{Clarendon Laboratory, Department of Physics,
	University of Oxford, Parks Road, Oxford OX1 3PU, UK}
\affiliation{Institute for Quantum Materials and Technologies, Karlsruhe Institute of Technology, 76021 Karlsruhe, Germany}

\author{Shiv J. Singh}
\thanks{Current affiliation: Institute of High-Pressure Physics, Polish Academy of Sciences, Sokolowska 29/37, 01-142, Warsaw, Poland}
\affiliation{Clarendon Laboratory, Department of Physics,
University of Oxford, Parks Road, Oxford OX1 3PU, UK}

\author{Nick Bultinck}
\affiliation{Rudolf Peierls Centre for Theoretical Physics,
	University of Oxford, Parks Road, Oxford OX1 3PU, UK}

\author{Amalia I. Coldea}
\email[corresponding author: ]{amalia.coldea@physics.ox.ac.uk}
\affiliation{Clarendon Laboratory, Department of Physics,
University of Oxford, Parks Road, Oxford OX1 3PU, UK}

\begin{abstract}

Iron-chalcogenide superconductors display rich phenomena caused by orbital-dependent band shifts and electronic correlations.
Additionally, they are potential candidates for topological superconductivity due to the band inversion between the Fe $d$ bands and the chalcogen $p_z$ band.
Here we present a detailed study of the electronic structure of the nematic superconductors FeSe$_{1-x}$Te$_x$ ($0<x<0.4$) using angle-resolved photoemission spectroscopy to understand the role of orbital-dependent band shifts, electronic correlations and the chalcogen band.
We assess the changes in the effective masses using a three-band low energy model, and the band renormalization via comparison with DFT band structure calculations.
The effective masses decrease for all three-hole bands inside the nematic phase followed by a strong increase for the band with $d_{xy}$ orbital character.
Interestingly, this nearly-flat $d_{xy}$ band becomes more correlated as it shifts towards the Fermi level with increasing Te concentrations and as the second superconducting dome emerges.
Our findings suggests that the $d_{xy}$ hole band, which is very sensitive to the chalcogen height, could be involved in promoting an additional pairing channel and increasing the density of states to stabilize the second superconducting dome in FeSe$_{1-x}$Te$_x$.
This simultaneous shift of the $d_{xy}$ hole band and enhanced superconductivity is in contrast with FeSe$_{1-x}$S$_x$.
\end{abstract}
\date{\today}
\maketitle

\section{Introduction}

Iron-chalcogenide superconductors (FeCh) display a wide variety of complex electronic phases due to their multi-band nature, small Fermi-energies and large orbital-dependent electronic correlations \cite{Tamai2010, Yi2015, Coldea2021}. These stabilize rich phase diagrams that include electronic nematic and spin-density wave phases \cite{Reiss2020, Mukasa2021}, topological surface states \cite{Zhang2018} and bring the system in the proximity of a BCS-BEC transition \cite{Rinott2017}. The electronic nematic phase manifests as a spontaneous breaking
 of the rotational symmetry induced by orbital and momentum dependent band shifts
 \cite{Fernandes2014b, Fedorov2016}.
Studies under small amounts of applied strain along the symmetry breaking channel lead to a divergent
elastoresistivity close to the transition temperature, revealing its electronic origin \cite{Kuo2016, Bartlett2021, Ghini2021,Watson2015a}.

Although, nematic critical fluctuations have been suggested to enhance superconductivity \cite{Bohmer2022},
the nematic phase is usually accompanied by a spin-density wave, making it difficult to isolate which interaction is driving the superconductivity
 \cite{Fernandes2022,Shibauchi2014, Gu2017, Malinowski2020}.
Moreover, the finite coupling with the lattice can be detrimental on
 the critical nematic fluctuations and superconductivity  \cite{Reiss2020,Labat2017}.
FeSe is a unique system which harbours an extended and tunable nematic electronic phase in the absence of long range magnetic order, whilst still stablizing superconductivity at low temperatures \cite{Coldea2018}.
 Studies of the superconducting gap have suggested a $s_\pm$ sign-changing pairing symmetry
 which is consistent with a spin-fluctuation mechanism  \cite{Sprau2016,Kreisel2020, Rhodes2018},
but competing Néel and stripe magnetic fluctuations are found in FeSe \cite{Wang2016a,Kreisel2020}.
In the presence of non-magnetic disorder both superconductivity and nematicity
is suppressed in Fe$_{1-x}$Cu$_x$Se studies \cite{Zajicek2022}.

The nematic order of FeSe can be effectively tuned via isoelectronic substitution of Se with S or Te, and
the band structure is very sensitive to the chalcogen height above the conducting planes  \cite{Watson2015a, Kumar2012}.
In addition, the Te substitution causes the chalcogen $p_z$ band to
shift down towards the hole pockets of the Fe $d$ bands and induce a band inversion at the Fermi level \cite{Wang2015},
which could stabilize topological surface states \cite{Lauke2020, Zhang2018, Chen2020, Fernandes2022}.
The orbital dependence of the electronic correlations becomes significant for high Te concentration
where  the band with $d_{xy}$ orbital character experiences a large increase in effective mass
relative to other bands with $d_{xz}/d_{yz}$ orbital character \cite{Tamai2010, Liu2015, Huang2022}.

Single crystals inside the nematic phase of FeSe$_{1-x}$Te$_x$ series with low Te content have only recently become available,
as the chemical vapour transport techniques have overcome the phase separation \cite{Xing2021, Ovchenkov2021}.
 Recent studies have determined the superconducting phase diagram
which harbours two superconducting regimes, one dome inside the nematic phase followed by a second dome which
 emerges at the nematic end point, proposed to be driven by critical nematic fluctuations \cite{Ishida2022}.

In this paper, we report a detailed ARPES study of the nematic FeSe$_{1-x}$Te$_x$ single crystals and determine
the evolution of the low-energy features
and the role of electronic correlations. Employing a detailed multi-band model, we extract the effective masses of all three
hole bands which decrease with increasing Te concentration as both the nematic order and superconducting phases are suppressed.
We find that
the effective mass of the hole band with $d_{xy}$ character displays
a non-monotonic variation reaching a minimum before increasing
significantly with the Te substitution.
Simultaneously, the $d_{xy}$ hole band shifts towards the Fermi level,
which could open a new pairing channel for superconductivity and an enhancement in the density of states,
potentially stabilizing a  second superconducting dome.
The $k_z$ dependence of the ARPES spectra
enables us to assess the contribution of the chalcogen $p_z$ band,
 which is instrumental in the formation of topological superconductivity.
Lastly, we explore the evolution of the electron pockets with Te concentration and find that size of the pockets increase,
whereas the nematic splitting slowly reduces as the nematic phase has a extended compositional range.

\begin{figure*}[htbp]
    \centering
 \includegraphics[width=\linewidth]{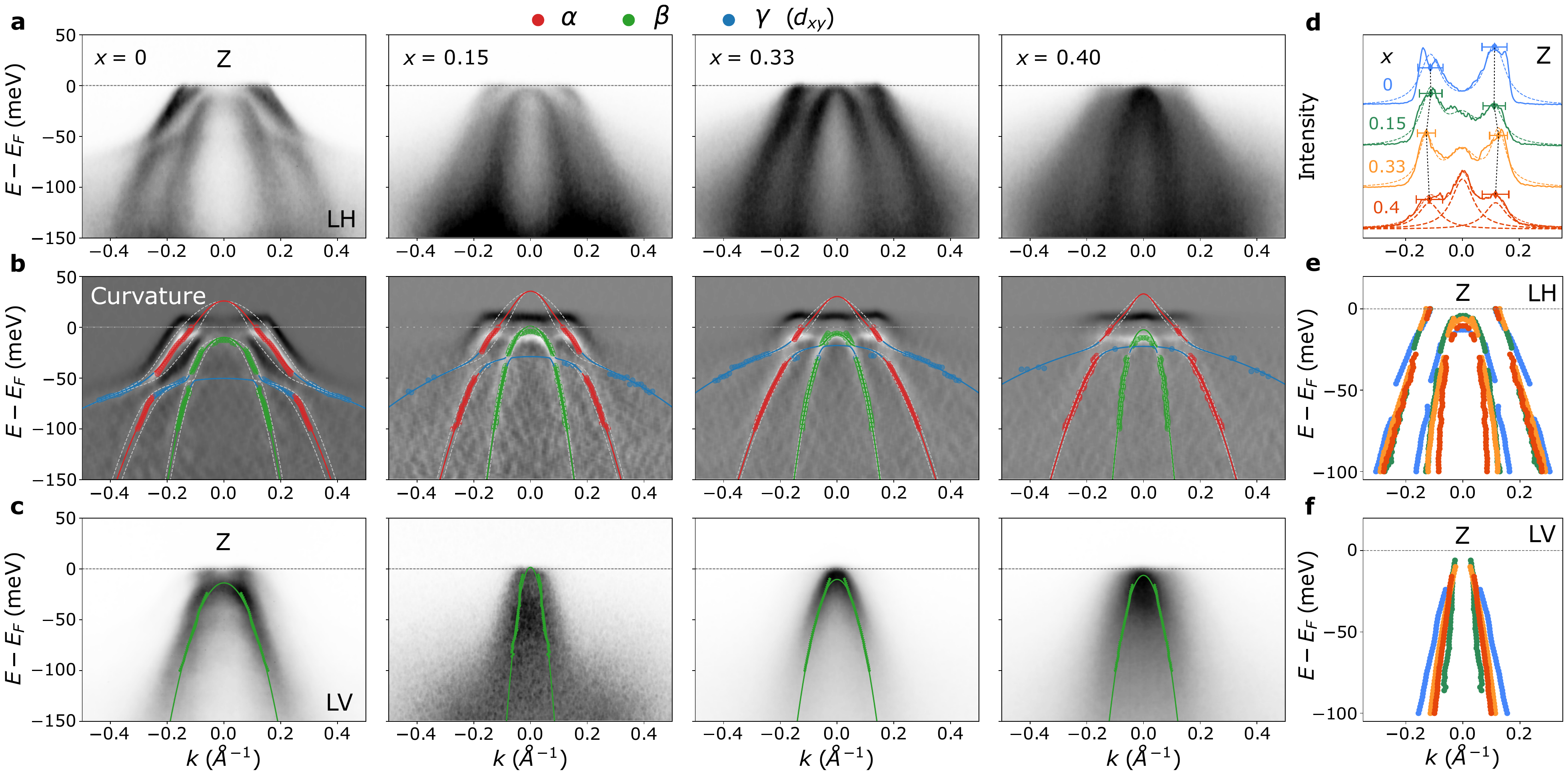}
    \caption{\textbf{Hole bands of nematic FeSe$_{1-x}$Te$_x$}.
    \textbf{(a)} ARPES spectra measured at the Z  high-symmetry point along the $A- Z - A$ direction at 10~K using linear horizontal (LH) light polarisation.
    \textbf{(b)} The corresponding two-dimensional curvature of the ARPES data from \textbf{(a)} using the method in Ref.~\cite{Zhang2011}.
    The solid points are the band positions extracted from Lorentzian fits to the
    MDCs ($\alpha$ and $\beta$ bands), maximums in the EDCs ($\beta$ band) and extrema in the curvature of the EDCs ($\gamma$ band).
       The solid lines are fits to the band positions using the low-energy model discussed in the SM.
    \textbf{(c)} ARPES spectra for the same dispersions as in \textbf{(a)} but measured using LV light polarisation.
    The solid points are the band positions extracted from MDC analysis, and the solid curve is a parabolic fit to the data.
    The solid lines in \textbf{(b)} and \textbf{(c)} corresponds to the outer hole band, $\alpha$ (red), the inner hole band, $\beta$ (green) and the flatter $d_{xy}$ hole band, $\gamma$ (blue).
    \textbf{(d)} MDCs taken at the Fermi level to determine the Fermi wavevector ($k_{\rm F}$) from \textbf{(a)}, where the errors are equal to half of the FWHM from the Lorentzian fits (dashed curves).
\textbf{(e)} The extracted hole dispersions of the $\alpha$ and $\beta$ bands  for different $x$ using LH polarization. \textbf{(f)} The extracted $\beta$ bands using LV polarization.
   Different colours  in \textbf{(d)}, \textbf{(e)} and \textbf{(f)} refer to the $x$ compositions, as indicated in \textbf{(d)}.
    }
    \label{Fig1}
\end{figure*}

\section{Results}

\subsection{The evolution of the hole bands}

The band structure of tetragonal FeSe, which is dominated by the bands with Fe 3$d$ orbital character at the Fermi level ($E_{\rm F}$),
consists of multiple hole pockets in the centre of the Brillouin zone
and electron pockets in the corners of it.
Fig.~\ref{Fig1}(a) shows the measured hole band dispersions along the $A - Z - A$ direction for different compositions of FeSe$_{1-x}$Te$_x$.
To highlight weaker spectral features for some of the bands,
we have calculated the corresponding curvature \cite{Zhang2011}, as illustrated in Fig.~\ref{Fig1}(b).
We clearly observe two dispersive bands with mixed $d_{xz/yz}$ orbital character,
 labelled as the $\alpha$ band for the outer hole band and the $\beta$ band for the inner band.
A third much flatter band with dominant $d_{xy}$ orbital character, labelled $\gamma$, has a weaker spectral weight and is seen $\sim 50$~meV below the Fermi level in FeSe \cite{Watson2015a}.
As the intensity of ARPES data is proportional to the matrix elements, which are dependent on the properties of both the incident light and the excited electrons,
we also use the linear vertical (LV) polarisation to isolate the $\beta$ hole band, as shown in Fig.~\ref{Fig1}(c) \cite{Zhang2012}.
Moreover, matrix elements suppress significantly the intensity associated with bands of $d_{xy}$ orbital character in both polarisations.
Therefore, the $\gamma$ band is mainly visible via hybridization with the $\alpha$ and $\beta$ bands and it can be visualized using the curvature analysis in Fig.~\ref{Fig1}(b).

Next, we discuss the band dispersions of the $\alpha$ and $\beta$ hole bands shown in Fig.~\ref{Fig1}(b).
To quantify precisely their position we use both momentum distribution
curves (MDCs) (fitted to pairs of symmetric Lorentzian peaks in Fig.~\ref{Fig1}(d))
and the maximum in the energy distribution curves (EDCs)  to define the top of the $\beta$ band
which is located mainly below $E_{\rm F}$.
We observe that the $\alpha$ hole band clearly crosses the Fermi level (see Fig.~\ref{Fig1}(c)),
whereas the spectral weight of the $\beta$ band only emerges from $x \sim 0.15$ as a central peak in the MDC spectrum at $E_{\rm F}$, as shown in Fig.~\ref{Fig1}(d).
In tetragonal FeSe the $\beta$ band forms a 3D hole pocket, however upon entering the nematic phase this pocket is pushed
below the Fermi-level \cite{Watson2015b,Coldea2021}. By suppressing the nematic phase with S substitution, the $\beta$ pocket re-emerges at FeSe$_{0.89}$S$_{0.11}$ \cite{Watson2015b,Coldea2021}
However, with Te substitution,
in which the nematic phase extends over a larger compositional range towards $x \sim 0.45$ \cite{Ishida2022},
the shift of this band is less pronounced.
Interestingly, the $\gamma$ band shifts significantly from a binding energy around 50~meV to  20~meV towards the Fermi level at both high symmetry points (see Figs.~\ref{Fig1}(b) and S6 in the Supplementary Material (SM))
This is in contrast to findings for the FeSe$_{1-x}$S$_x$ series, where the $\gamma$  band is relatively unaffected by isoelectronic substitution up to $x \sim 0.18$ \cite{Reiss2017}.
However, these findings are in good agreement with the prediction of the DFT calculations which shows that the $\gamma$ band only begins to shift for value of $h >1.45$ \AA (see Fig.~S3(d) in the SM).

Fig.~\ref{Fig1}(e) and (f) show that the extracted band dispersions for the $\alpha$ and $\beta$ hole bands become narrower, suggesting a decrease in the effective masses with increasing Te concentration.
The extracted values of the Fermi wavevector, $k_F$, suggest that  the size of the $\alpha$ hole pocket hardly varies with increasing Te concentration,
consistent with studies at higher $x$ values \cite{Ieki2014} (see Figs.~\ref{Fig1}(d) and ~S7 in the SM)

The nematic order parameter, $\varphi_{nem}$ is usually estimated from the energy splitting between the $\alpha$ and $\beta$ hole bands,
given by $\Delta_h= \sqrt {\varphi^2_{nem}+ \lambda^2_{1}}$ \cite{Fernandes2014a,Xing2017}.
As the  spin-orbit coupling (SOC) parameter,  $\lambda_{1}\sim 23.5$~meV for FeSe at high temperatures
inside the tetragonal phase \cite{Watson2015a, Watson2017b,Coldea2018}, and at low temperatures $\Delta_h$ increase towards $37.5$~meV, the value of $\varphi_{nem}$ was estimated as $\sim$~29meV for FeSe \cite{Watson2017b,Coldea2018}.
With S isoelectronic substitution $\varphi_{nem}$ decreases and at low temperatures the $\beta$ hole band crosses the Fermi level and forms the 3D pocket at $Z$ point \cite{Watson2015b,Reiss2017,Coldea2021}.
In contrast, the $\Delta_h$ splitting for FeSe$_{1-x}$Te$_x$ remains around 38(3)~meV due to two competing effects:
one caused by the reduction in the nematic order parameter with increasing $x$ and the second caused by the enhanced SOC splitting
(DFT calculations predict a variation from 82~meV for FeSe to 177~meV for FeTe at the $Z$ point, as shown in Fig.~S1 in the SM).

\subsection{The role of the chalcogen $p_z$ band}

The position of the $p_z$ band and its interaction with the hole bands is complex and heavily dependent on the chalcogen height, $h$, which strongly changes with the Te concentration, $x$,
  as shown in Fig.~S1 and S2 in the SM.
The chalcogen $p_z$ band,  which is predicted to be located above the Fermi level at the $\Gamma$ point in FeSe,
intercepts vertically all three hole bands along the $\Gamma-Z$ direction (see $E_{p_z}$ in Fig.~S1 in the SM) .
Due to symmetry reasons, the crossing between the inner hole band and the chalcogen $p_z$ band is proposed to create a band inversion,
allowing topological surface states to form \cite{Zhang2018, Fernandes2022}.
Based on the DFT predictions,  this crossing first occurs at the $Z$ point for $h \approx 1.4$~\AA, and its hybridisation with the inner hole band creates a gap (along the $Z-A$ direction).
As a consequence the $p_z$ band and the inner hole band smoothly merge and their orbital characters mix, resulting in a pseudo-inner hole band having a larger degree of $p_z$ orbital character \cite{Wang2015}.

In order to assess the contribution of the $p_z$ band in experiments,
we perform a quantitative analysis of the ARPES intensity of the $\beta$ band in relation to the $\alpha$ band, $I(\beta,\alpha)$. This relative intensity is evaluated at the two high-symmetry points with different $k_z$ values ($\Gamma$ and $Z$) giving the ratio $I(\beta,\alpha)_Z/I(\beta,\alpha)_\Gamma$.
We find this ratio to increase beyond $x=0.33$, as shown in Figs.~\ref{Fig1} and S7 in the SM.
This variation in intensity indicates an alteration in the orbital character of the bands due to hybridization with other bands. However, changes in intensity due to nematic effects and the contribution from the $\gamma$ band should manifest equally at the $\Gamma$ and $Z$ point.
Thus, the variations in the intensity ratio suggest that the matrix elements have changed for the $\beta$ band due to the interaction with the $p_z$ band, which is strongly $k_z$ dependent.
This is in agreement with previous studies which have highlighted the presence of the $p_z$ band at $Z$ in $x\sim 0.55$ \cite{Lohani2020, Wang2015,Li2023}.
Additionally, we observe the emergence of a central peak in the MDCs at $E_F$ for $x \sim 0.15$, which could also be due to the enhanced intensity of the $\beta$ band at $Z$.

\begin{figure*}
    \centering
    \includegraphics[width=\linewidth]{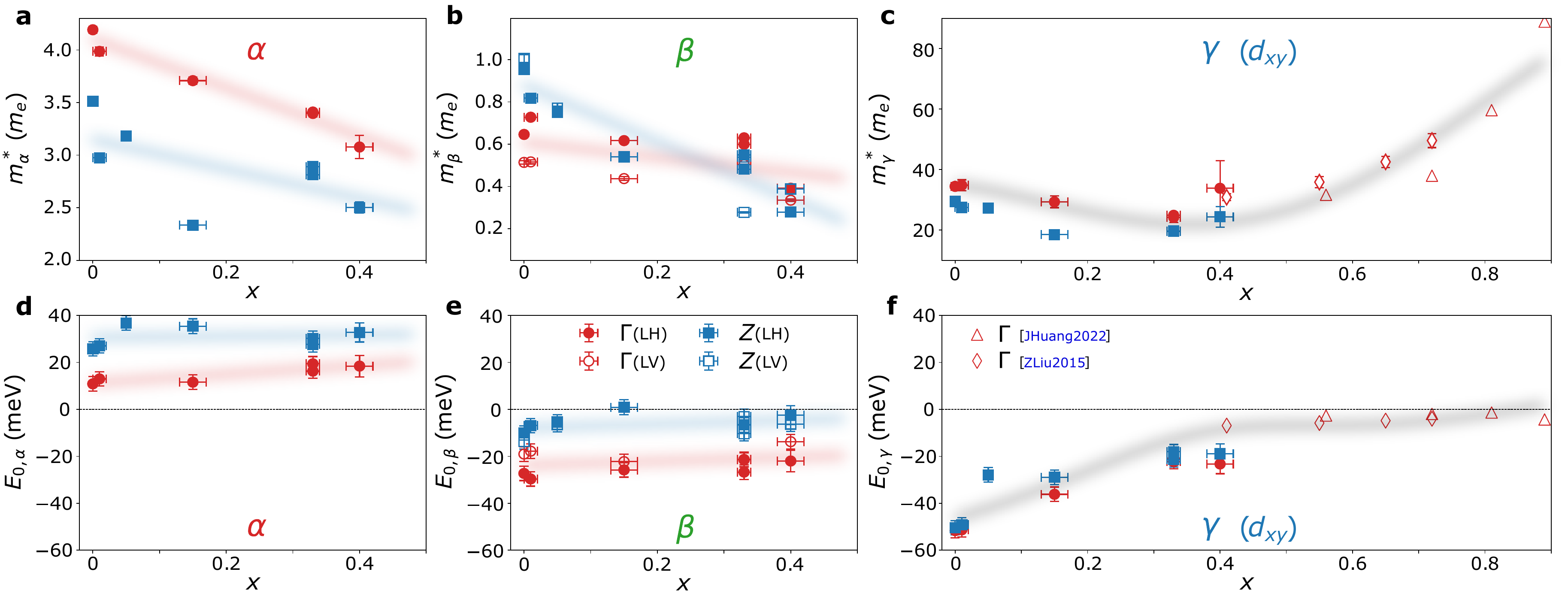}
    \caption{\textbf{Effective masses and Fermi energies of the hole bands}.
  The effective masses of the $\alpha$ band in \textbf{(a)}, the $\beta$ band in \textbf{(b)} and the $\gamma$ hole band in \textbf{(c)} as a function of Te concentration.
    The solid lines in \textbf{(a)} and \textbf{(b)} are linear fits  for each data set at different high symmetry points, $\Gamma$ (red circles) and $Z$ (blue squares).
     Data for $\beta$ are for both polarisations, with open and solid symbols corresponding to LV and LH, respectively.
    The solid line in \textbf{c} is a 5\textsuperscript{th} order polynomial fit to the data.
    Open diamond and triangles are from Refs.~\cite{Huang2022} and \cite{Liu2015}, respectively.
    The Fermi energies of each band are plotted as a function of $x$
     in \textbf{(d)} ($E_{0, \alpha}$), \textbf{(e)} ($E_{0, \beta}$) and \textbf{(f)} ($E_{0, \gamma}$), where data from Refs. \cite{Huang2022} and \cite{Liu2015} are added.
    The solid lines in \textbf{(d)} and \textbf{(e)} are linear fits, while the solid line in \textbf{(f)} is described by a phenomenological model detailed in the text.
    }
    \label{Fig2}
\end{figure*}

\subsection{The low-energy model and effective masses}

In order to quantify the evolution of the electronic correlations and band shifts of the hole bands of FeSe$_{1-x}$Te$_x$ as a function of Te concentration,
we employed a low energy three-band model to describe the effective masses  of the $\alpha$, $\beta$ and $\gamma$ bands.
This model contains essentially three parabolic dispersions in the presence of SOC and the nematic order.
This approach is an extension of the model developed from Refs.~\cite{Cvetkovic2013, Fernandes2014a, Rhodes2021} to accommodate the third hole band, $\gamma$,
as detailed in Fig. S4 in the SM.
To constrain the parameters of the model we fixed $\varphi_{nem}$ to be proportional to the nematic transition temperature, $T_s$, such that $\varphi_{nem} = 29$~meV for FeSe,
as reported in Ref.~\cite{Watson2015a,Watson2017b}.
This additional constraint was introduced to account for the fact that the nematic order
parameter is difficult to estimate from the
 splitting between the $\alpha$ and $\beta$ hole bands, $\Delta_h$.
 This is caused by the lack of experimental data above the Fermi level and linewidth broadening caused by the impurity scattering and averaging over two different domains,
as detailed in the SM. 

Fig.~\ref{Fig2} shows the extracted effective masses, $m^*$, from the three-band model as a function of Te concentration.
For the $\alpha$ and $\beta$ hole bands, $m^*$ decreases with increasing Te substitution. This is similar to the renormalisation factor for the $d_{z^2}$ band, as shown in Fig.~S8(c) in the SM.
In contrast, the effective mass of the $\gamma$ hole band displays a non-monotonic behaviour, as it decreases inside the nematic phase
(from $x=0$ towards a minimum close to $x\sim 0.3$) and then increases to values twice as large as those seen in FeSe at $x\sim 0.8$  \cite{Liu2015, Huang2022}.

The renormalisation factors for each band can be estimated from the ratio of the experimental effective masses to the band masses evaluated from DFT calculations.
The estimated values for  FeSe are 3.5 and 1.8 for the $\alpha$ and $\beta$ band, respectively,
which is in close agreement to those previously reported \cite{Maletz2014}. However, for the $\gamma$ band we determine a value $\sim 21$,
whereas previous studies give values between 8 and 10 \cite{Watson2015a, Maletz2014}.
This discrepancy lies within our use of the curvature analysis, which gives us a larger momentum range to constrain
 the fit and our consideration for the SOC between the $\gamma$ and $\alpha$ bands.

\subsection{Evolution of the electron pockets}

\begin{figure*}[htbp]
    \centering
     \includegraphics[width=\linewidth]{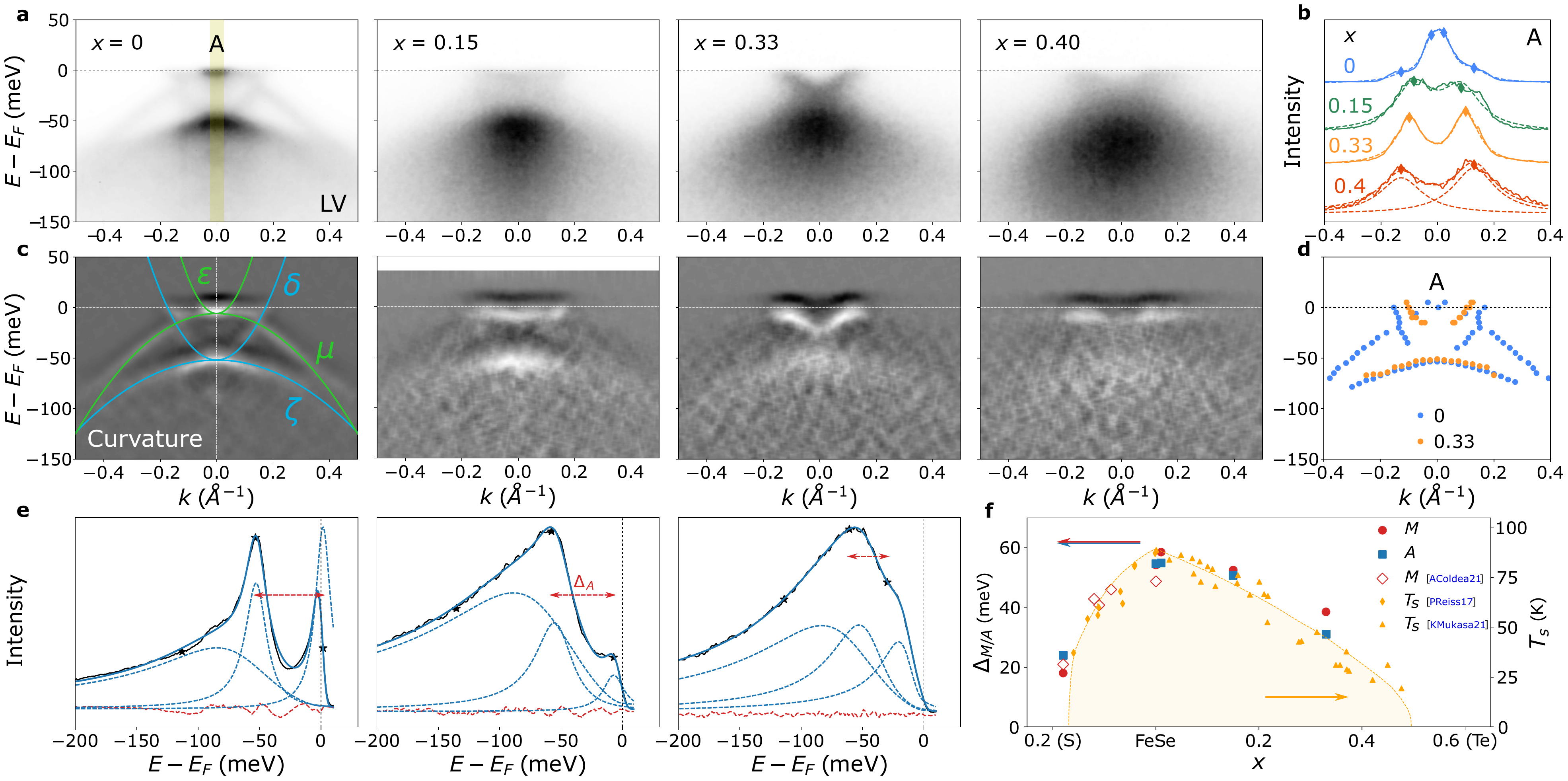}
    \caption{\textbf{Electron Pocket of FeSe$_{1-x}$Te$_x$.}
    \textbf{(a)} ARPES spectra taken at 10~K along the $Z-A-Z$ direction using LV light polarisation for the different compositions, $x$.
    \textbf{(b)} MDCs  at the Fermi level for each $x$ and the dashed plots correspond to Lorentzian fits. The diamond data points indicate the Lorentzian centers, and thus the $k_{\rm F}$ values.
    \textbf{(c)} The two-dimensional curvature of the ARPES spectra in \textbf{(a)}. The different parabolic dispersions are  labelled for the FeSe spectra.
    \textbf{(d)} The estimated band dispersion for $x=0$ and $x=0.33$, where the data is extracted from the EDC minima in the curvature plots.
    \textbf{(e)} The estimation of $\Delta_A$ by fitting the EDC data (solid black curve) (integrated within the window indicated by the yellow strip for $x=0$ in \textbf{(a)}).
     The blue solid curve is the fit, the dashed blue curves are the individual Lorentzian curves and dashed red plot is the residuals. Star data points correspond to the position of the Lorentzian centre.
     \textbf{(f)}  The variation of $\Delta_{A}$ as a function of composition, $x$, (solid square) together with those for $\Delta_M$ (solid circles) from Fig.~S11 in the SM.
     Additional data for FeSe$_{1-x}$S$_x$ (open red rhombuses) are taken from Ref.~\cite{Coldea2021}. The variation of $T_{\rm s}$ as a function of composition is plotted on right $y$-axis using data from Refs.~\cite{Reiss2017} (orange diamond) and \cite{Mukasa2021} (orange triangle).
    }
    \label{Fig4}
\end{figure*}

Fig.~\ref{Fig4}(a) shows the evolution of the ARPES spectra corresponding
to the electron pockets centred at the $A$ high-symmetry point for different Te substitutions.
We observe significant changes between the spectra for the different compositions, as compared
with FeSe, in particular at high Te concentration ($x\sim 0.33$) where only one large electron pocket is detected.
The DFT calculations in the tetragonal phase indicate that there are two degenerate points
($A_1$ corresponds to two bands with $d_{xz/yz}$ character
whereas $A_3$ is for two bands of $d_{xy}$ character)
\cite{Fernandes2014a} (see Fig.~S1 in the SM).
The energy separation between spectral features along the EDC at the $A$ point, defined as $\Delta_{A}$,
 is finite inside the tetragonal phase and found to be $\sim 20$~meV for FeSe$_{0.82}$S$_{0.18}$ \cite{Coldea2021}.
However, as the nematic phase emerges the degeneracy of the two points are lifted (see Fig.~S10 in the SM)
the value of $\Delta_{A}$ has been originally interpreted as a parameter to quantify the nematic order \cite{Yi2019, Zhang2016, Watson2016},
 increasing towards $\sim 50$~meV \cite{Watson2015a} in bulk FeSe or $\sim 70$~meV in thin films of FeSe \cite{Zhang2016}.

 Fig.~\ref{Fig4}(e)  illustrates the estimation of $\Delta_{A}$ for different FeSe$_{1-x}$Te$_x$
from the EDC curves, as the separation between the two upper-most Lorentzian centres where the energy broadening is assumed to be linearly dependent on energy.
 This linear energy dependence, often associated to a marginal Fermi-liquid behaviour in iron-based superconductors \cite{Fink2021},
 is evaluated by analysing the $\beta$ hole band dispersion,
  as shown in Fig.~\ref{Fig1}(c) and Fig.~S6(c) in the SM.
Fig.~\ref{Fig4}(f) shows that there is a direct correlation between $\Delta_{A}$ and the structural transition $T_{\rm s}$,
similar to FeSe$_{1-x}$S$_x$ \cite{Coldea2021},  which decreases as the nematic order is suppressed with increasing Te substitution.
 However, to assess the precise nematic electronic order parameter
any additional splitting would need to be quantified for single domain detwinned samples under applied strain, as in the case of FeSe
\cite{Fernandes2014a,Christensen2020, Rhodes2021, Watson2016, Yi2019, Zhang2016, Fedorov2016}.

Fig.~\ref{Fig4}(d) shows a comparison between the different band dispersions centered at the $A$ point,
using the extremal values from the curvature plots due to their weak features.
Interestingly, we find the position of the hole-like band dispersion, $\zeta$,
remains unchanged  up to $x \sim 0.33$, as shown in Fig.~\ref{Fig4}(c), where this continues even for higher Te concentrations ($0.56<x<0.75$) \cite{Liu2015}.
Furthermore, only one electron pocket is resolved for higher $x$
(using the MDC curves in Fig.~\ref{Fig4}(b))  and the $k_{\rm F}$ values increase with  Te concentration.
This is possibly due to the electron pockets becoming more isotropic as the nematic phase is suppressed.
Additionally, the Fermi surfaces map of $x=0.33$ (see Fig.~\ref{Fig5}(e)) shows that the
size of the electrons pockets relative to the hole
pockets is larger, suggesting that the system may be slightly electron doped (the average $k_F$ values for $x$ = 0.4 is 0.12~\AA$^{-1}$ for the electron pockets and  0.09~\AA$^{-1}$ for the hole pockets).

\section{Discussion}

\begin{figure}
    \centering
  \includegraphics[width=\linewidth]{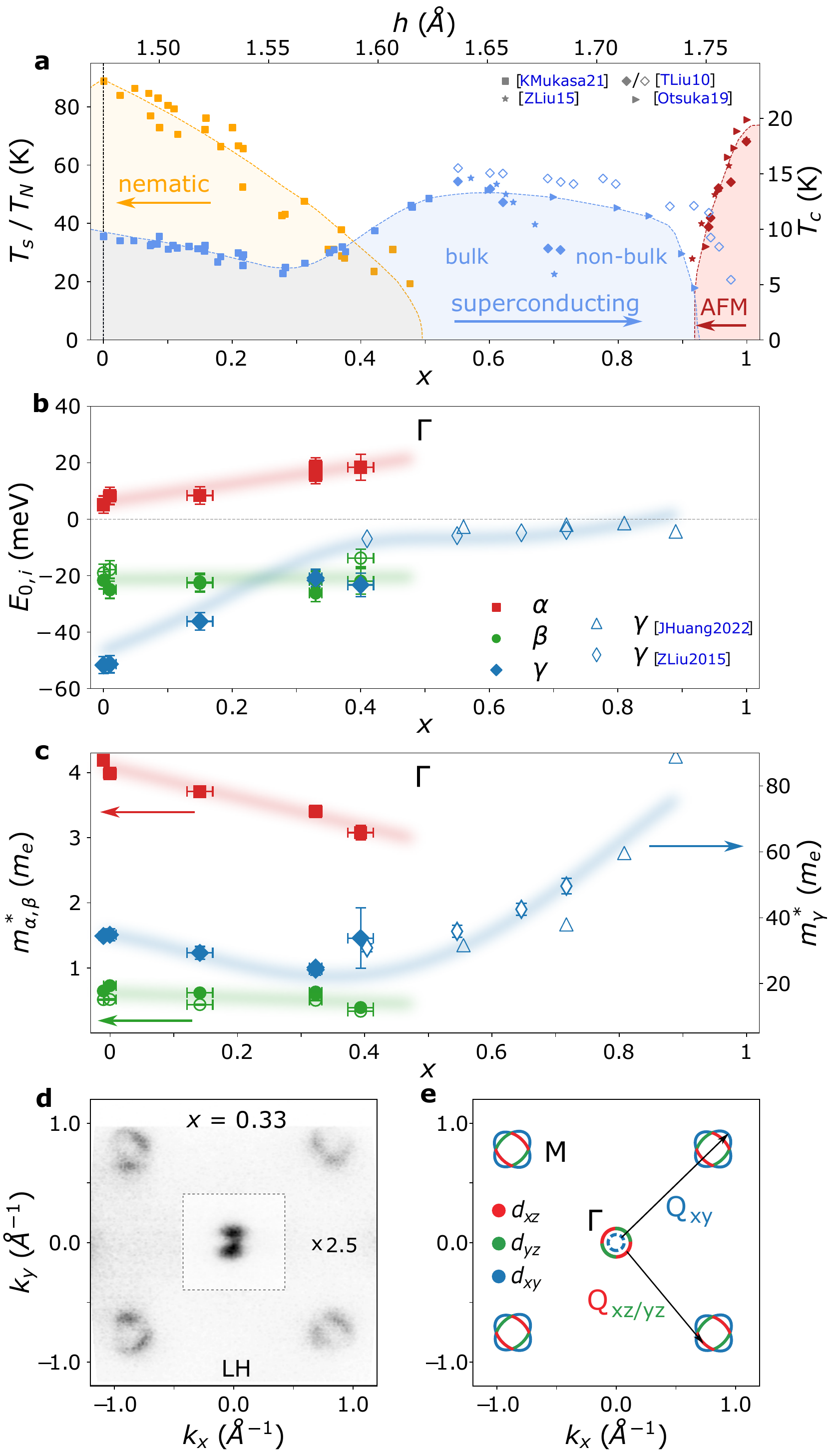}
    \caption{\textbf{(a)} Phase diagram of FeSe$_{1-x}$Te$_x$ which indicates the nematic phase transition at $T_{\rm s}$
    (orange squares), the two-dome superconductivity described by the critical temperature, $T_{\rm c}$
    (blue points) and the bi-collinear antiferromagnetic phase given by the variation of $T_N$ (red points).
    Data points are taken from Refs.~\cite{Mukasa2021}, \cite{Liu2010}, \cite{Liu2015} and \cite{Otsuka2019}.
    \textbf{(b)} The Fermi energies as a function of composition, $x$ for the $\alpha$ band (solid red squares),
     $\beta$ band (solid (LH) and open (LV) green circles) and  $\gamma$ band
     (solid blue rhombuses). Additional points for $\gamma$ band  are from Refs.~\cite{Huang2022} (blue triangles) and \cite{Liu2015} (diamonds).
    The solid red and green lines are linear fits to the $\alpha$ and $\beta$ band data, while the blue curve is the 5\textsuperscript{th} order polynomial fit to the $\gamma$ data.
    \textbf{(c)} Effective masses as a function of composition, $x$, where the symbols are defined as in \textbf{(b)}. The solid red and green lines are linear fits, while the blue curve is the phenomenological fit discussed in the main text.
    In \textbf{(b)} and \textbf{(c)} only data extracted at the $\Gamma$ point are shown.
    \textbf{(d)} Fermi surface map of the first Brillouin zone for $x=0.33$ taken at 10~K using LH polarisation.
    The spectra outside of the dashed box has been enhanced by a factor of 2.5.
    \textbf{(e)} Schematic Fermi surface indicating the dominant orbital character of each potion of the bands and potential pairing channels.
    }
    \label{Fig5}
\end{figure}

The phase diagram of FeSe$_{1-x}$Te$_x$ illustrates the suppression of the nematic phase with increasing
the Te concentration, $x$,
as shown in Fig.~\ref{Fig5}(a).
The superconducting transition temperature, $T_c$, decreases inside the nematic phase towards a minimum of $\sim6$~K around $x \sim 0.3$, followed by an increase towards 13~K above $x \sim 0.5$.
This resurgent superconducting dome extends over a large range and becomes filamentary
as the system approaches the bi-collinear magnetic order close to FeTe \cite{Liu2010}.
Fig.~\ref{Fig5}(c) shows  that the effective masses, $m^*$ of the $\alpha$, $\beta$ and $\gamma$ hole bands
 decrease by increasing $x$, which correlates with the reduction of the renormalisation factor associated the $d_{z^2}$ hole band, as shown in Fig.~S8(c) in the SM.
Interestingly, the reduction of the effective mass of the $\alpha$ hole band, which crosses the Fermi level, is accompanied by the suppression
 of superconductivity inside the nematic phase, suggesting this hole pocket is likely to be involved in the superconducting pairing.
 This is consistent with previous trends detected from quantum oscillations in FeSe$_{1-x}$S$_x$, where both the effective mass
 and the superconducting transition temperature, $T_{\rm c}$, reach a maximum near $x \sim 0.1$, followed by a strong suppression towards FeS
  \cite{Coldea2019,Coldea2021,Reiss2017},
 as well as for FeSe$_{0.89}$S$_{0.11}$ under pressure  \cite{Reiss2020}.

The changes in the effective masses also reflect the potential role played by the chalcogen $p_z$ band.
The rate of decrease of the effective mass, $m^*$, for the $\alpha$ band with the Te concentration, $x$,
 is very similar  for both high symmetry points.
 On the other hand,  the effective mass for the lighter $\beta$ band decreases faster at the $Z$ point, as compared with the $\Gamma$ point (Fig.~\ref{Fig2}(b)).
When the $p_z$ band intersect the $\beta$ band the orbital mixing between the two is strongest at the Z point.
As the $p_z$ orbital is assumed to be less correlated than the Fe $d$ orbitals \cite{Lohani2020},
it is expected that the effective mass of the $\beta$ band should
be further reduced at the $Z$ point relative to the $\Gamma$ point, in agreement with the observed experimental trends.

Fig.~\ref{Fig5}(b) and Fig.~\ref{Fig2}(d-f) show the Fermi energies, $E_0$, of the different bands  estimated from the low-energy model as a function of $x$,
indicating that $\alpha$ and $\beta$ bands experience minimal variation.
On the other hand, the $\gamma$ band with the $d_{xy}$ orbital character varies significantly and shifts towards the Fermi level
up to $x \sim 0.4$, then it plateaus in its vicinity.
Experimentally, there is a degree of uncertainty whether the $\gamma$ band crosses the Fermi level
 \cite{Tamai2010, Ieki2014, Rinott2017} or not \cite{Liu2015, Huang2022}, and this discrepancy could be due to the amount of excess Fe  in these systems \cite{Rinott2017}.
Regardless, this is the region in the phase diagram where the second superconducting dome is stabilized, suggesting that the hole band
with $d_{xy}$ orbital character could be involved in pairing.
Furthermore, the strong shift  of the $\gamma$ band towards the Fermi level and the accompanying enhancement
of superconductivity was also seen in thin films of FeSe$_{1-x}$Te$_x$/CaF$_2$ \cite{Nakayama2021}.

To understand the significant shift towards the Fermi level
of the $\gamma$ band one can consider two mechanisms.
Firstly, DFT predicts that the $\gamma$ band moves up linearly with Te concentration, $x$,
as the chalcogen height, $h$, increases and the lattice expands (see Figs.~\ref{Fig5}(b) and Fig.~S1 in the SM).
Secondly, variations of the bandwidth, due to electronic correlations, will also cause the Fermi energy to change, which can be crudely scaled as $\sim 1/m^*$.
Thus, by combining the effect of the lattice and of the electronic correlations, the changes in the position of the $\gamma$ band
could be described by a phenomenological model, $ E_\gamma(x) = \nu_1 x + \nu_2 / m^*_\gamma(x) + \nu_3 $.
Fig.~\ref{Fig5}(c) shows a fit to the experimental positions of the $\gamma$ band, where $\nu_1$ is the weight of the lattice effect, $\nu_2$ is the weight of the bandwidth and $\nu_3$ is an arbitrary offset ($m^*_\gamma(x)$ is taken as a 5\textsuperscript{th} order polynomial fit from Fig.~\ref{Fig2}(c)).
We find the fit captures the initial increase followed by a plateau for $x>0.4$.
For even higher $x$ values  a new hole-like pocket with $d_{z^2}$ orbital character was detected
 at the $X$ point for $x \geq 0.8$ due to the increase in electronic correlations which leads to the loss
of the spectral weight of the $d_{xy}$ band \cite{Huang2022}.

The presence of the third hole band with $d_{xy}$ orbital close to
the Fermi level and its involvement in pairing could manifest in different ways.
Firstly, the hybridisation of the $\alpha$ hole band with the $\gamma$ band would lead to a reduction in the Fermi velocity, thus increasing the density of states
and enhancing superconductivity.
Secondly, the orbital mixing between these bands
would open up a new pairing channel which connects to the $d_{xy}$ orbital character at the electron pocket with the hole pockets (see Fig.~\ref{Fig5}(e)).
Thirdly, a new pairing channel could be achieved if $E_{0,\gamma}$ is smaller than the superconducting gap, as
in this regime the $\gamma$ band can play the role of an incipient band \cite{Chen2015}.
The superconducting pairing in  FeSe is proposed to be mediated by spin fluctuations between the hole and electron pockets
with $d_{yz}$ orbital character along $Q_{xy/yz}=(\pi,\pi)$ (Fig.~\ref{Fig5}(e)) \cite{Rhodes2018, Marzin2008, Wang2016a, Gu2017}.
A similar vector, $Q_{xy}$, that would connect the $d_{xy}$ channel and the observation of stripe magnetic
fluctuations in FeSe$_{1-x}$Te$_x$ could suggest that the second superconducting dome may also be driven by spin fluctuations \cite{Li2021, Argyriou2010, Liu2010}.
However, neutron scattering studies also observe the development of ($\pi,0$) spin fluctuations which compete with superconductivity \cite{Liu2010}.
This competition is evident as the suppression of superconductivity, towards FeTe, is accompanied by the development of a double-stripe magnetic
order with a $(\pi,0)$ wave vector \cite{Otsuka2019}.

There is a strong contrast between FeSe$_{1−x}$Te$_x$
and FeSe$_{1−x}$S$_x$ series, which displays no enhancement of $T_{\rm c}$ outside the nematic phase
and the $d_{xy}$ hole band is located 50~meV below the Fermi level \cite{Coldea2021}.
Additionally, quantum oscillations found a direct correlation between the evolution of the effective mass
of the $\alpha$ band and $T_{\rm c}$ in FeSe$_{1−x}$S$_x$ \cite{Coldea2019}, which also occurs inside the nematic phase of FeSe$_{1−x}$Te$_x$.
These trends change significantly in the vicinity of  the nematic end point of FeSe$_{1−x}$Te$_x$  ($x \sim 0.5$),
as  the effective masses for the $\alpha$ and $\beta$ hole bands remain rather constant  
 \cite{Liu2015, Huang2022}, whereas the superconductivity is enhanced (see Fig.~\ref{Fig5}(a)).
 On the other hand,  both the effective mass for the $\gamma$ band and $T_{\rm c}$  increase
at a similar composition $x$. This suggests that the mechanism for superconductivity likely changes
 across the phase diagram  of  FeSe$_{1-x}$Te$_x$ \cite{Liu2015, Yin2011} and
 the hole band $\gamma$, with $d_{xy}$ orbital character, plays an important role.

An alternative mechanism for the enhancement of superconductivity can be related to its proximity to the nematic end point.
Theoretically, nematic fluctuations should support superconductivity in both an $s$ or $d$-wave pairing channel \cite{Labat2017, Tamai2010},
and elastoresistivity studies find a divergence of the nematic susceptibility near its end point \cite{Ishida2022}.
However, studies of FeSe$_{0.89}$S$_{0.11}$ under pressure suggest the critical fluctuations could
 either be quenched via the coupling of the nematic order to the lattice \cite{Reiss2020} or obscured within the nematic islands of a quantum Griffiths phase \cite{Reiss2021}.
Moreover, other hallmarks of a quantum critical point in FeSe$_{1-x}$Te$_x$, such as the divergence of the effective masses
 at the nematic end point and/or a linear dependence of resistivity in temperature are not observed \cite{Mukasa2021}.
These systems are rather bad metals, with large linewidths in the ARPES spectra,
which could be caused by the large distribution of the Se and Te ions (chalcogen heights differ by 0.24~\AA)  \cite{Tegel2010}
and the stabilization of the interstitial Fe that could act as additional scattering centers.

The evolution of the band structure and the significant shifts of the $d_{xy}$ band via Te substitution
have implications on other studies.
The chalcogen height $h$ is an important parameter that influences the changes in
the band structure, magnetic ground state and superconductivity \cite{Kumar2012, Kuroki2009}.
For example,  $h$, could be shifted towards higher values with applied pressure in other iron chalcogenides, despite the shrinking of the unit cell.
Theoretical studies of FeSe and NMR studies of FeSe$_{0.88}$S$_{0.12}$  have suggested that the hole $d_{xy}$ band
 could lead to the enhanced superconducting dome induced by pressure \cite{Yamakawa2017, Kuwayama2021}.
However, pressure also induces magnetic order and the maximum in $T_{\rm c}$ is often located at the end point of the magnetic
phase in FeSe$_{1-x}$S$_x$ \cite{Matsuura2017, Mukasa2021},
suggesting that magnetic fluctuations are important for superconducting pairing.

\section{Conclusion}

In this study we present a comprehensive investigation of the low-temperature band structure of the nematic FeSe$_{1-x}$Te$_x$ series.
We identify a direct correlation between the changes in the nematic transition  with the
spectral features in the dispersions at the corner of the Brillouin zone, in agreement with other studies
on iron-chalcogenide superconductors. Additionally, the changes in the relative intensities of spectra corresponding
to the inner and outer hole bands reflect the contribution of the $p_z$ band, as
it is brought towards the Fermi level by increasing Te concentration.

By using a low-energy three-band model, which captures much of the underlying electronic properties,
we extract the effective mass for the hole bands that show an unexpected decrease inside the nematic phase.
These findings are consistent with the decrease in the renormalisation associated the higher energy $d_{z^2}$ bands.
On the other hand, the effective mass of the third hole band with $d_{xy}$ orbital character
 has a local minimum before increasing significantly in the tetragonal region.
 Concomitantly, we detect the gradual shift of this, rather flat, $d_{xy}$ hole band towards the Fermi-level with increasing Te concentration,
  which could lead to an enhancement of the density of states and the opening of a new pairing channel.
This effect could be responsible for the emergence of the second superconducting dome outside the nematic phase
involving the hole band with $d_{xy}$ orbital character, in contrast to the
superconducting dome inside the nematic phase.
These trends occur at the same time as the chalcogen height increases,
and our findings may have implications for other studies, like those under applied pressure.
Moreover, at high Te concentration, the resistivity of FeSe$_{1-x}$Te$_x$ reflects
 bad metallic behaviour but superconductivity remains rather robust.
Our findings suggest changes in the superconducting mechanism between that emerging
from the nematic phase, with rather weak electronic correlations,  and the second
superconducting dome outside the nematic phase.
Further theoretical studies that aim to understand the superconducting mechanisms in these systems
could consider the presence of the orbitally dependent electronic correlations, the
enhanced density of states induced by a nearly-flat band with $d_{xy }$ orbital character
and potentially the significant effects caused by the local variation in the chalcogen height.

\vspace{0.5cm}

{\bf Methods}
Single crystals of FeSe$_{1-x}$Te$_x$ were grown via the new chemical vapour transport method \cite{Terao2019} and screened via transport studies and X-ray diffraction (XRD).
The nematic transition, $T_{\rm s}$ was evaluated from the temperature dependence of the longitudinal resistivity
(as the extremum in the first derivative as a function of temperature) as well as from X-ray studies, and compared with a previous report \cite{Mukasa2021}.
After each ARPES experiment, we also performed Energy Dispersive X-ray (EDX) spectroscopy studies to
 directly measure the composition of each investigated crystal and these values are reported here.

ARPES measurements were performed at the I05 beamline at the Diamond Light Source synchrotron using a MBS A1 hemispherical analyser giving a combined energy resolution of $\sim$3~meV \cite{Hoesch2017}.
The incident photon energy was varied between 23-100~eV to probe the $k_z$ dependence of the electronic structure and to identify the high-symmetry points.
The single crystals were cleaved {\it in-situ} using a top-post and all measurements were performed at the base temperature of 10~K and under ultrahigh vacuum at $10^{-10}$mbar. The beamspot was $\sim50 \mu$m which is much larger than the nematic domains of the samples \cite{Rhodes2020}. Therefore, all spectra are a superposition of the two orthogonal domains. Besides the reported data on new crystals of FeSe$_{1-x}$Te$_x$,  we compare our results to previous ARPES data on FeSe from Refs.~\cite{Watson2015a,Coldea2018}
and the tetragonal system, FeSe$_{0.82}$S$_{0.18}$, reported previously in Refs.~\cite{Reiss2017,Coldea2021}.

To complement the experimental data,
we have performed DFT band structure calculations using the Wien2k software in the GGA approximation with and without SOC \cite{Wien2k} which are  presented in the SM.
These calculations were performed in the tetragonal phase with lattice parameters taken from Refs.~\cite{Reiss2017, Mukasa2021, Li2009},
 as listed in Table~1 in the SM.

\vspace{0.5cm}

{\bf Acknowledgements}
We thank Luke Rhodes for useful discussions on the low-energy model.
ABM acknowledge scholarship funding from the Department of Physics,
via OxPEG and the Diamond Light Source.
We thank Diamond Light
Source for access to Beamline I05 (Proposals No. SI30564, SI29118)
that contributed to the results presented here.
This work was mainly supported by
 EPSRC (Grants No. EP/I004475/1 and No. EP/I017836/1)
 and the Oxford Centre for Applied Superconductivity.
A.A.H. acknowledges the financial support of the Oxford Quantum Materials
Platform Grant (EP/M020517/1).
 A.A.H. acknowledges the financial support of the Deutsche Forschungsgemeinschaft
 (DFG, German Research Foundation) TRR288–422213477 (project B03).
 The authors would like to acknowledge the use of the University
of Oxford Advanced Research Computing (ARC) facility in
carrying out part of this work \cite{ARC}. A.I.C. acknowledges support
from an EPSRC Career Acceleration Fellowship (Grant No.
EP/I004475/1).

\vspace{0.5cm}

{\bf Data availability}
In accordance with the EPSRC policy framework on
research data, access to the data will be made available from
the Oxford University Research Archive
ORA (https://ora.ox.ac.uk/) using the following link http://dx.doi.org/10.5287/ora-e9k5bn9e8.

{\bf Author contributions}
A.B.M., T.K, M.D.W, A.I.C performed
ARPES measurements.
A.AH., S.J.S grew the crystals;
A.B.M analyzed the ARPES data;
A.B.M., T.K, M.D.W, A.I.C  discussed the results and manuscript preparation.
A.B.M and A.I. C wrote the manuscript with contributions from
all authors.

{\bf Competing interests}
The authors declare no competing interests.

{\bf Additional information}
Supplementary information is available for this paper.

\bibliography{refs}

\clearpage
\newpage

\onecolumngrid

\section{Supplemental Material}

\onecolumngrid

\newcommand{\blue}{\textcolor{blue}}
\newcommand{\bdm}[1]{\mbox{\boldmath $#1$}}

\renewcommand{\thefigure}{S\arabic{figure}} 
\renewcommand{\thetable}{S\arabic{table}} 

\newlength{\figwidth}
\figwidth=2\textwidth

\setcounter{figure}{0}

\newcommand{\fig}[3]
{
    \begin{figure}[!tb]
        \vspace*{-0.1cm}
        \[
        \includegraphics[width=\figwidth]{#1}
        \]
        \vskip -0.2cm
        \caption{\label{#2}
            \small#3
        }
\end{figure}}

\subsection*{DFT calculations}
\label{Appendix: DFT}

To be able to compare the ARPES experiments and understand the evolution of the band structure in FeSe$_{1-x}$Te$_x$, we have performed detailed density functional theory (DFT) calculations in the tetragonal structure for two separate series. We use the GGA approximation as implemented in Wien2k \cite{Wien2k} on the ARC server at the University of Oxford \cite{ARC}.

Firstly, DFT calculations were performed using the experimental parameters, as listed in Supplementary Table~\ref{Lat Params}.
 The input parameters in the DFT calculations of FeSe$_{1-x}$Te$_x$ assume that the chalcogen site
  is occupied by Se  and to account for different $x$ compositions the lattice parameters
  and the chalcogen height  are modified  (after Ref.~\cite{Mukasa2021}),
 whereas for FeTe the structure is taken from Ref.~\cite{Li2009}.
Due to the isoelectronic substitution of a larger Te ion for Se, the lattice  parameter $c$ and the value of $h$ varies significantly
 whereas the $a$ lattice parameter hardly changes in FeSe$_{1-x}$Te$_x$.
We find that the $d_{xy}$ band shifts to lower binding energy as the unit cell expands and it has little $k_z$ dependence, suggesting a quasi-two dimensional pocket, as shown in Supplementary Figure.~\ref{SupFig_DFT}.
The two $d_{z^2}$ bands also shift  as their average position move away from $E_F$, but their separation also increases,
whereas the degenerate point of the $d_{xz/yz}$ bands  (as indicated by $E_D$ in Supplementary Figure~\ref{SupFig_DFT}a),  hardly changes by increasing $x$.
However, the $p_z$ band is highly sensitive to Te substitution and it has a strong $k_z$ dependence.
It is predicted to cross the Fermi level in FeSe and shift to higher binding energies with increasing $x$,
whereas the $k_z$ dependence of the $p_z$ band becomes stronger for FeTe, suggesting the development of stronger interlayer coupling and a warping of the Fermi surface.

Secondly, further calculations were performed for FeSe whereby only the chalcogen height, $h$, was varied (see Supplementary Figure~\ref{SupFig_DFT_ch}).
By varying the values of $h$, it reveals a strong shift of the $p_z$ band.
The full series nicely illustrates the hybridisation of the $p_z$ band and the inner $d_{xz/yz}$ hole band, as the $p_z$ band shifts from being above $E_F$ to below with increasing $h$.
At low $h$ values ($h = 1.382 $\AA), the $p_z$ is isolated from the hole bands creating an electron-like dispersion. However, as $h$ increases, the band shifts and crosses these hole bands ($h \approx 1.4$~\AA), where it hybridises strongly with the inner hole band (or $\beta$ band).
Along the $Z-A$ direction, this creates a gap such that the $p_z$ band appears to smoothly merge with the inner hole band forming the pseudo-$\beta$ band.
Moreover, we also observe the $d_{xy}$ band to shift to lower binding energies, but the degenerate point ($E_D$) remains relatively constant as $h$ increases.
Therefore, the effect of tuning $h$ is rather similar to the isoelectronic substitution discussed above.

\begin{table}[h]
\caption{Lattice parameters used for the DFT calculations in the tetragonal symmetry.
 $a$ and $c$ are the tetragonal lattice parameters and $h=z \cdot c$ is the height of the chalcogen above
the Fe plane. The system crystallises into the P4/nmm space group (no. 129).
}
\centering
\begin{tabular}{cccccc}
\hline
  \hline
  FeSe$_{1-x}$Te$_x$ & chalcogen & $a$ (\AA) & $c$ (\AA) & $h$ (\AA) & $z$  \\
  \hline
  $x=0$ & Se & 3.7651 & 5.5178 & 1.4744 & 0.2672 \cite{Reiss2017} \\
  $x=0.15$  & Se & 3.7766 & 5.6435 & 1.5144 & 0.2683 \cite{Mukasa2021} \\
  $x=0.33$  & Se & 3.7862 & 5.8120 & 1.5743 & 0.2709 \cite{Mukasa2021} \\
  $x=0.40$  & Se & 3.7900 & 5.8775 & 1.5975 & 0.2718 \cite{Mukasa2021} \\
  $x=1$ & Te & 3.8120 & 6.2517 & 1.7686 & 0.2791 \cite{Li2009} \\
  \hline
  \hline
\end{tabular}
\label{Lat Params}
\end{table}

\begin{figure}[h]
    \centering
     \includegraphics[width=0.8\linewidth]{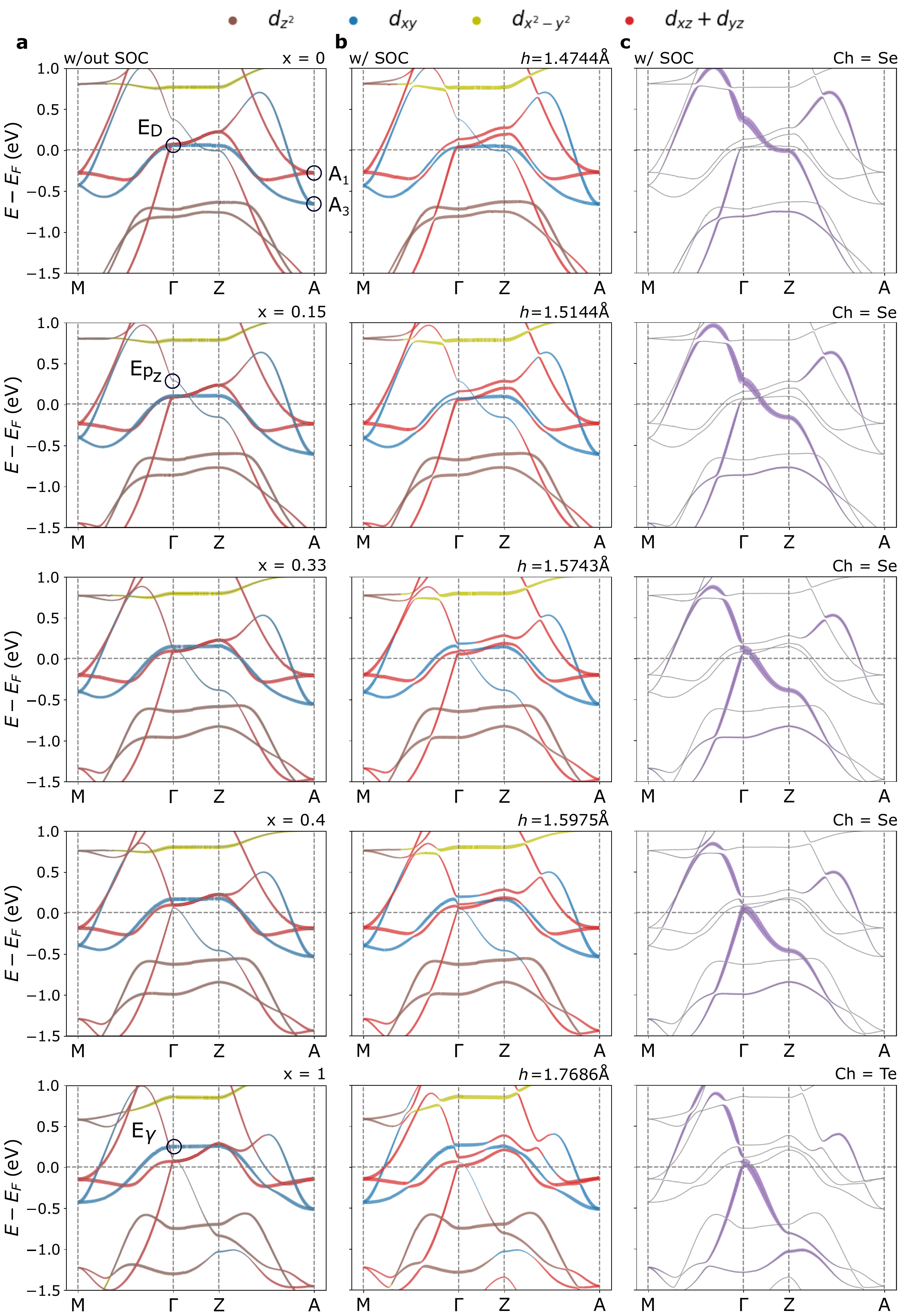}
    \caption{\textbf{DFT band structure calculations of  FeSe$_{1-x}$Te$_x$.}
    The calculated band dispersion in the tetragonal phase across the $M-\Gamma-Z-A$ path in $\vec{k}$-space without spin-orbit coupling (SOC) in  \textbf{(a)} and with spin-orbit coupling in  \textbf{(b)}.
      The different rows correspond to the different compositions, where the value of $x$, $h$ and the species of chalcogen atom are given and listed in Supplementary Table.~\ref{Lat Params}.
    In \textbf{(a)} and \textbf{(b)} only the Fe $d$ orbitals are considered, whereas in \textbf{(c)} only the contribution of the chalcogen $p_z$ orbital is displayed.
    The colours of the bands indicate the dominant orbital character (the $d_{xz}$ and $d_{yz}$ orbitals are degenerate in the $C_4$ symmerty)
    and the linewidth indicates the weight of that orbital. In \textbf{(c)} the linewidths have been scaled up by a factor of 6 relative to \textbf{(a)} and \textbf{(b)}.
    }
    \label{SupFig_DFT}
\end{figure}

To quantitatively compare the DFT with experiments we extract the band masses and Fermi-energies of each band, as shown in Supplementary Figure~\ref{SupFig_DFT_params}.
To extract these parameters we use the calculations in the absence of the spin-orbit coupling (see Supplementary Figure~\ref{SupFig_DFT}a) and each band was fitted to a parabola.
The band mass variation of the $\gamma$ band from FeSe to FeTe is small and remains close to 2$m_e$,
whereas for the $\alpha$ band, the band mass is similar  but increases; more so at the $\Gamma$ point.
However, for the $\beta$ band, the band mass is very light (below 0.5~$m_e$) and slightly decreases.
similar to the trend found for the experimental effective mass  (see Fig.~2).
 Moreover, band mass values of the $\beta$ band at $Z$ point are excluded here due to the strong hybridisation with the chalcogen $p_z$ band.

Supplementary Figure~\ref{SupFig_DFT_params} shows a comparison between the extracted values of the band mass and the Fermi energy of each band,
 as determined from the first DFT series, corresponding to the Te substitution, and the second series, in which only $h$ is varied.
 We observe that the two DFT series produce very similar results with the exception of the band mass of the $\alpha$ band.
 Supplementary Figure~\ref{SupFig_DFT_params}b shows that the $\alpha$ band can become extremely flat at high values of $h$, whereas the variation from FeSe to FeTe is much smaller.
Moreover, another surprising result was the $\gamma$ band does not shift for $h <1.45$ \AA~ and only for values above this,
 suggesting that FeSe is close to a threshold point where the $d_{xy}$ orbitals of Fe no longer strongly interact with the chalcogen atoms.

\begin{figure}[h]
    \centering
        \includegraphics[width=\linewidth]{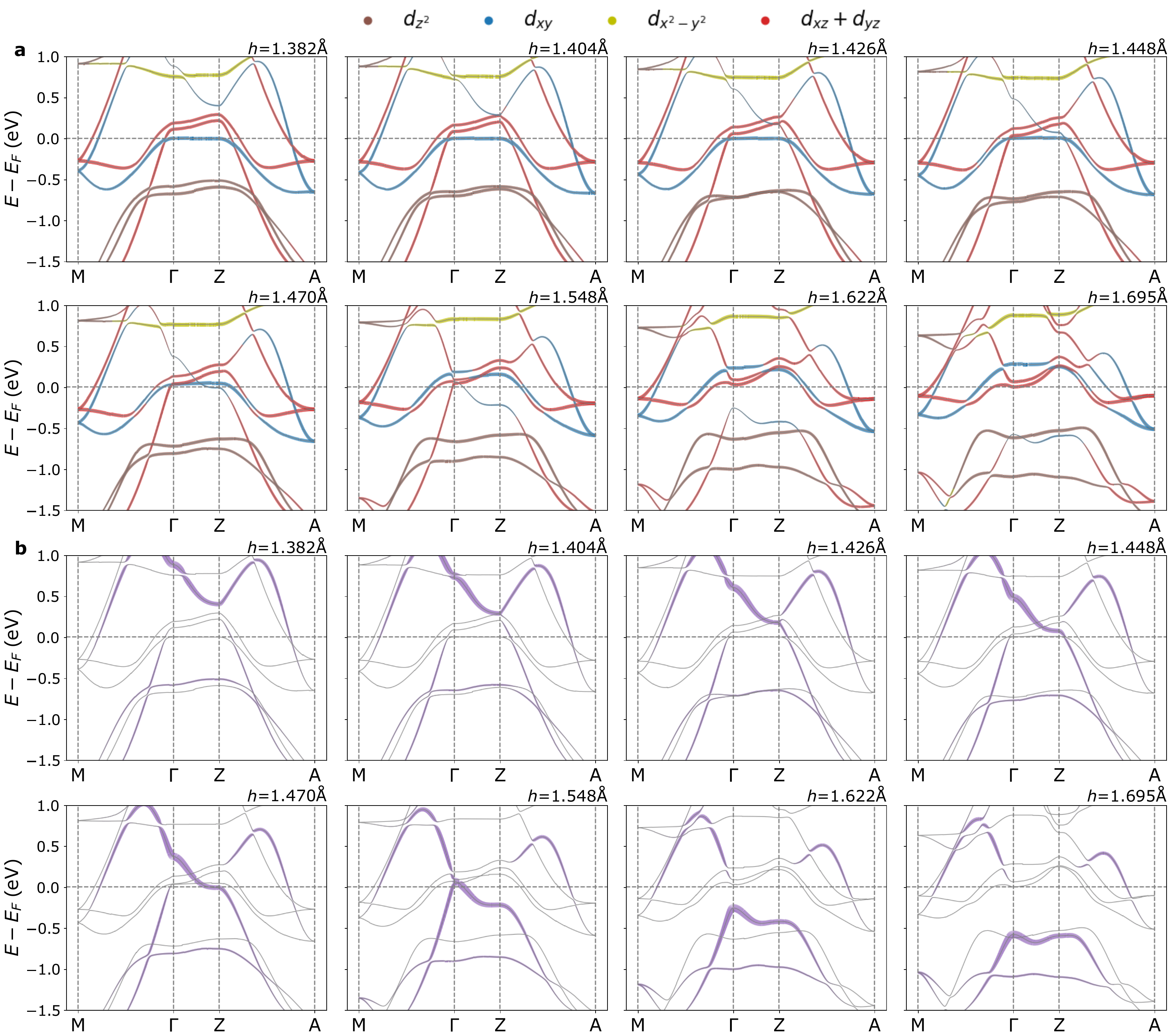}
    \caption{\textbf{Band structure calculations of FeSe by varying only the chalcogen height.}
     All calculations are performed in the tetragonal phase in the presence of the spin-orbit coupling along the $M-\Gamma-Z-A$ path in the $\vec{k}$-space.
     \textbf{(a)} Different panels show the evolution of the band structure as $h$ increases from 1.382~\AA ~to 1.695~\AA ,
      where the band colour and width indicate the dominant orbital character and its corresponding weight.
      In \textbf{(a)} only the Fe $d$ orbital character is displayed, whereas in \textbf{(b)}
       only the chalcogen $p_z$ orbital character is plotted. The line widths in \textbf{(b)} are scaled up by a factor of 6 in relation to \textbf{(a)}.
       }
    \label{SupFig_DFT_ch}
\end{figure}

\begin{figure}[h]
    \centering
        \includegraphics[width=\linewidth]{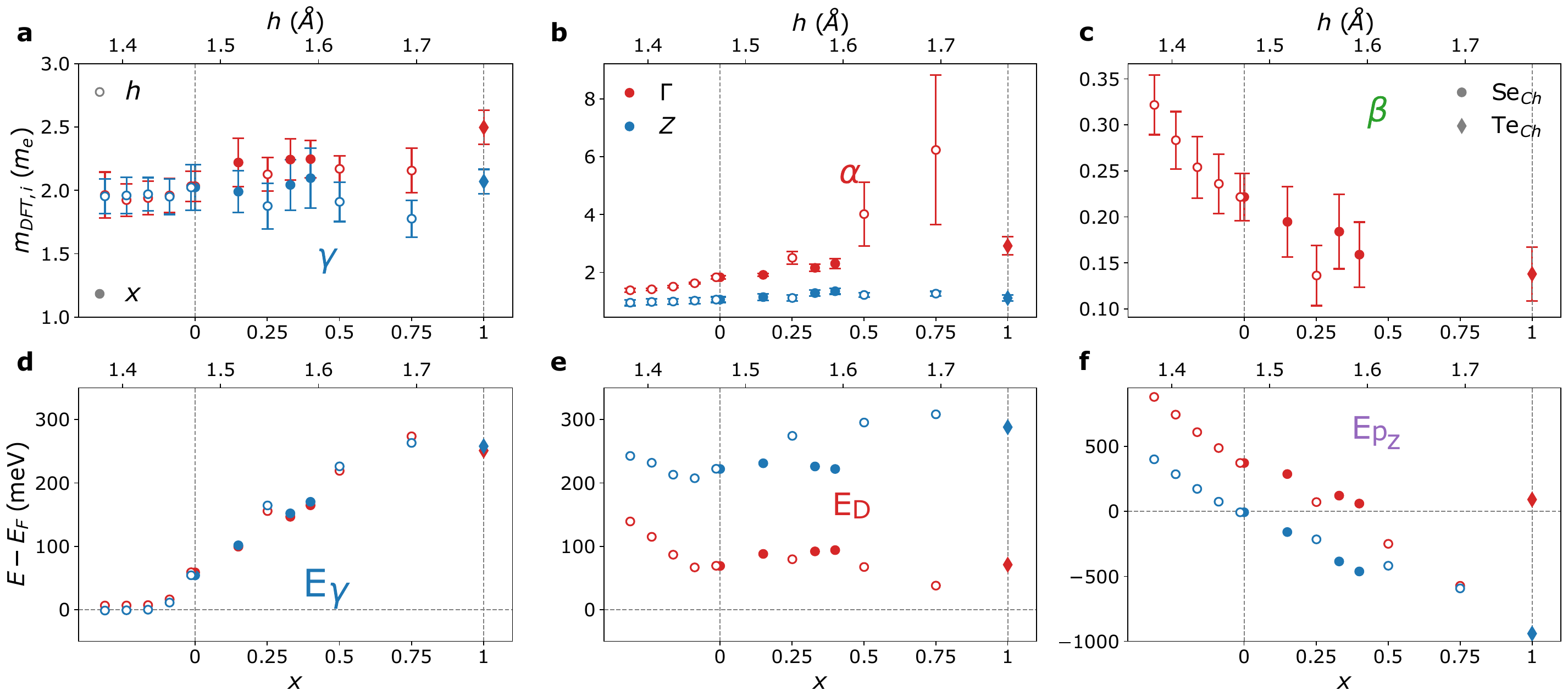}
    \caption{\textbf{Variation of the parameters extracted from the DFT calculations.}
      \textbf{(a)} Band mass of the $\gamma$ band with $d_{xy}$ character,
      \textbf{(b)} the outer band, $\alpha$, with $d_{xz/yz}$ character and \textbf{(c)} the inner band, $\beta$, with $d_{xz/yz}$ character. We observe hardly any variation in the calculated masses of the $\gamma$ band.
      The band masses of the $\alpha$ band increase, whereas those of the $\beta$ band decrease with $x$.
      \textbf{(d)} Fermi energy of the band with $d_{xy}$ character ($\gamma$), defined as $E_\gamma$ in Supplementary Figure~\ref{SupFig_DFT}(a).
      \textbf{(e)} Fermi energy of the bands with $d_{xz/yz}$ character (degenerate point of $\alpha$ and $\beta$) defined  as $E_D$ in Supplementary Figure~\ref{SupFig_DFT}(a).
      \textbf{(f)} Fermi energy of the chalcogen $p_z$ band defined as $E_{p_z}$ in Supplementary Figure~\ref{SupFig_DFT}(a).
      Each panel displays these values both as a function of composition, $x$ (solid symbols corresponding to the bottom axis) as well as $h$ values (open symbols corresponding to the top axis).
      Red and blue data points are for the $\Gamma$ and $Z$ points respectively. Circles data points signify that Se atoms were used for the calculation, whereas diamonds signify that Te was used.}
    \label{SupFig_DFT_params}
\end{figure}

\newpage
\clearpage

\subsection*{The low-energy three-band model used to describe the hole bands}
\label{Appendix: Model}

Here we present the three-band model used to parameterize the behaviour of the hole band dispersions at the high symmetry points $Z$ (Fig.1) 
and $\Gamma$ (Supplementary Figure~\ref{SupFig_Gamma}). This model consists of three-parabolic dispersions in the presence of both spin-orbit coupling and the nematic order. It is based on models previously discussed in Refs.~\cite{Cvetkovic2013, Fernandes2014a, Rhodes2021} and  expanded to accommodate the third hole band of $d_{xy}$ character, $\gamma$, as detailed in Supplementary Figure~\ref{SupFig_model}. The Hamiltonian to describe this three-band model is given by,

\begin{equation}
    H =
    \begin{pmatrix}
        c_{\alpha,\beta} + a k^2 + b(k_x^2-k_y^2) - \varphi_{nem}/2 & -2 b k_x k_y - i \lambda_1/2 & -i \lambda_2 k_x /\sqrt{2}\\
        -2 b k_x k_y + i \lambda_1/2 & c_{\alpha,\beta} + a' k^2 - b(k_x^2-k_y^2) + \varphi_{nem}/2 & i \lambda_2 k_y /\sqrt{2}\\
        i \lambda_2 k_x /\sqrt{2} & -i \lambda_2 k_y /\sqrt{2} & c_\gamma + a_\gamma k^2\\
    \end{pmatrix}
    \longrightarrow
    \begin{pmatrix}
        E_{yz} \\
        E_{xz} \\
        E_{xy}
    \end{pmatrix}
\label{Ham}
\end{equation}

which conserves the point group symmetry around both the $\Gamma$ and $Z$ high symmetry points. Here $\lambda_1$ corresponds to the SOC between the bands of $d_{xz}$ and $d_{yz}$ character ($\alpha$ and $\beta$) and $\lambda_2$ relates to the SOC of the bands of $d_{yz/xz}$ character and the third additional band of $d_{xy}$ character (the $\gamma$ band). One should note that $\lambda_2$ has the units of velocity as the SOC is momentum dependent. Moreover, $\varphi_{nem}$ is the nematic order parameter, which reflects the change in symmetry from C4 to C2.
Supplementary Figure~\ref{SupFig_model} illustrates the effect of different
parameters on the low-energy model, where SOC and the nematic order are added in stages.

\begin{figure}[htbp]
    \centering
       \includegraphics[width=\linewidth]{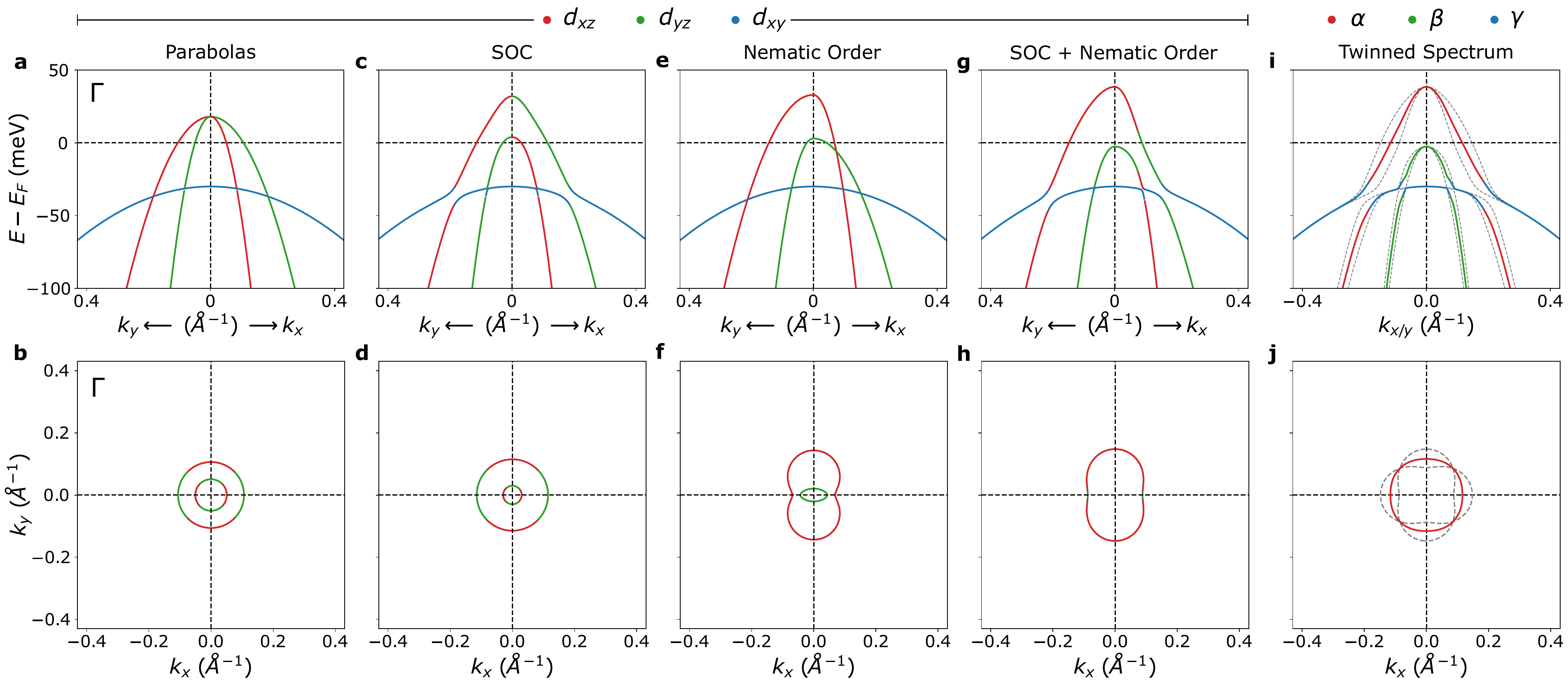}
    \caption{\textbf{The low-energy model and the evolution
    of the  band  dispersions and the Fermi surfaces.}
    The model uses the Hamiltonian given in equation~\ref{Ham} using the following parameters,
    $a = -4300$~meV$\cdot$\AA$^2$,
    $b = 2700$~meV$\cdot$\AA$^2$,
        $a_\gamma = -200$~meV$\cdot$\AA$^2$,
    $c_{\alpha,\beta} = 18$~meV and
    $a_\beta = -30$~meV.
    \textbf{(a, c, e, g, i)} Give the band dispersions of the model when the spin-orbit coupling (SOC) terms and nematic order parameters are varied, while \textbf{(b, d, f, h, j)} give the Fermi surfaces corresponding to the above band dispersion. The different columns correspond to the following assumptions:
    \textbf{(a, b)} Both the nematic order and the SOC are set to zero;
    \textbf{(c, d)} The SOC is finite ($\lambda_1=14$~meV and $\lambda_2=45$~meV$\cdot$\AA), but the nematic order is still zero;
    \textbf{(e, f)} The nematic order is finite ($\varphi_{nem}=30$~meV), but both SOC terms are zero;
    \textbf{(g, h)} Both the nematic order and SOC terms have finite values ($\varphi_{nem}=30$~meV, $\lambda_1=14$~meV and $\lambda_2=45$~meV$\cdot$\AA).
    The colours of the bands  in \textbf{(a-h)} correspond to the orbital content, where red, green and blue correspond to $d_{xz}$, $d_{yz}$ and $d_{xy}$ respectively.
    \textbf{(i, j)} The twinned model, where both the nematic order and SOC terms have finite values ($\varphi_{nem}=30$~meV, $\lambda_1=14$~meV and $\lambda_2=45$~meV$\cdot$\AA), however the dispersion along a certain direction is averaged with the dispersion at 90$^\circ$ to it (about the $k=0$ axis). In \textbf{(i)} the dispersions along $k_x$ and $k_y$ are plotted as dashed grey curves, giving the two domains, and the average (in energy) then forms the coloured plots. The same is true for the Fermi surface in \textbf{(j)} where the dashed grey plots give the two domains, while the coloured plots give the average between them. In \textbf{(i)} and \textbf{(j)}, the colours red, green and blue correspond to the $\alpha$, $\beta$ and $\gamma$ hole bands respectively.}
    \label{SupFig_model}
\end{figure}

\begin{figure}[htbp]
    \centering
        \includegraphics[width=\linewidth]{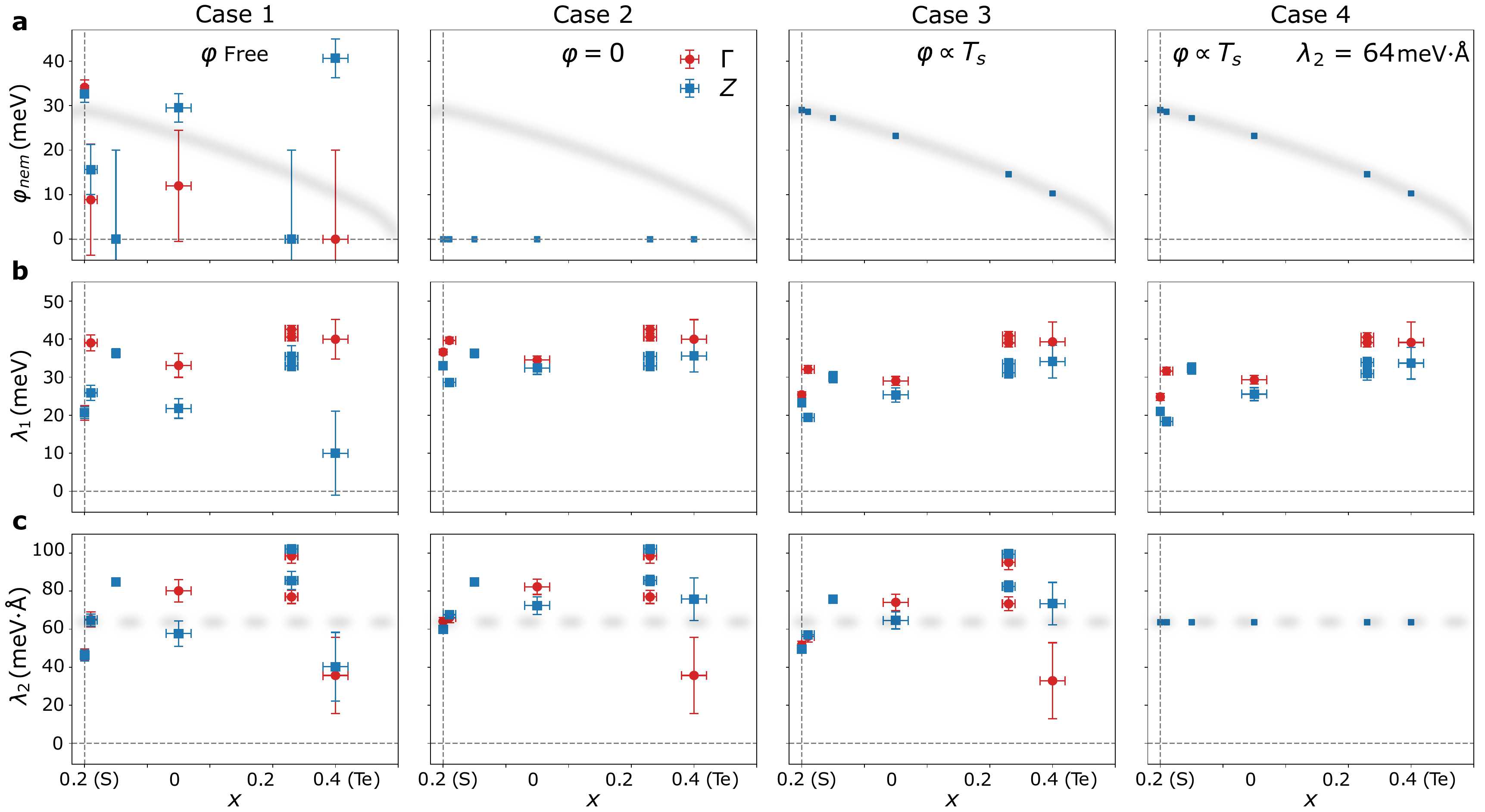}
    \caption{
    \textbf{Variation of different parameters in the low-energy model.}
    Different columns correspond to
    various cases related to the values of $\varphi_{nem}$ and
    the $\lambda_2$.
     In row \textbf{(a)} the variation of the nematic order parameter ($\varphi_{nem}$) is plotted as a function of Te concentration for the different cases.
    Rows \textbf{(b)} and \textbf{(c)} show the variation of $\lambda_1$ and $\lambda_2$, respectively, as a function of Te concentration. The broad, solid, grey curve in \textbf{(a)} indicates the value of $\varphi_{nem}$ when it is proportional to $T_{\rm s}$. The broad, dashed, grey curve in \textbf{(c)} indicates the value of $\lambda_2$ when it is set to 64~meV$\cdot$\AA.
        In all plots red circles correspond to data taken at the $\Gamma$ point, while blue square correspond to data taken at the $Z$ point.
    }
    \label{SupFig_Diff_Model}
\end{figure}

\begin{figure}[h]
    \centering
        \includegraphics[width=\linewidth]{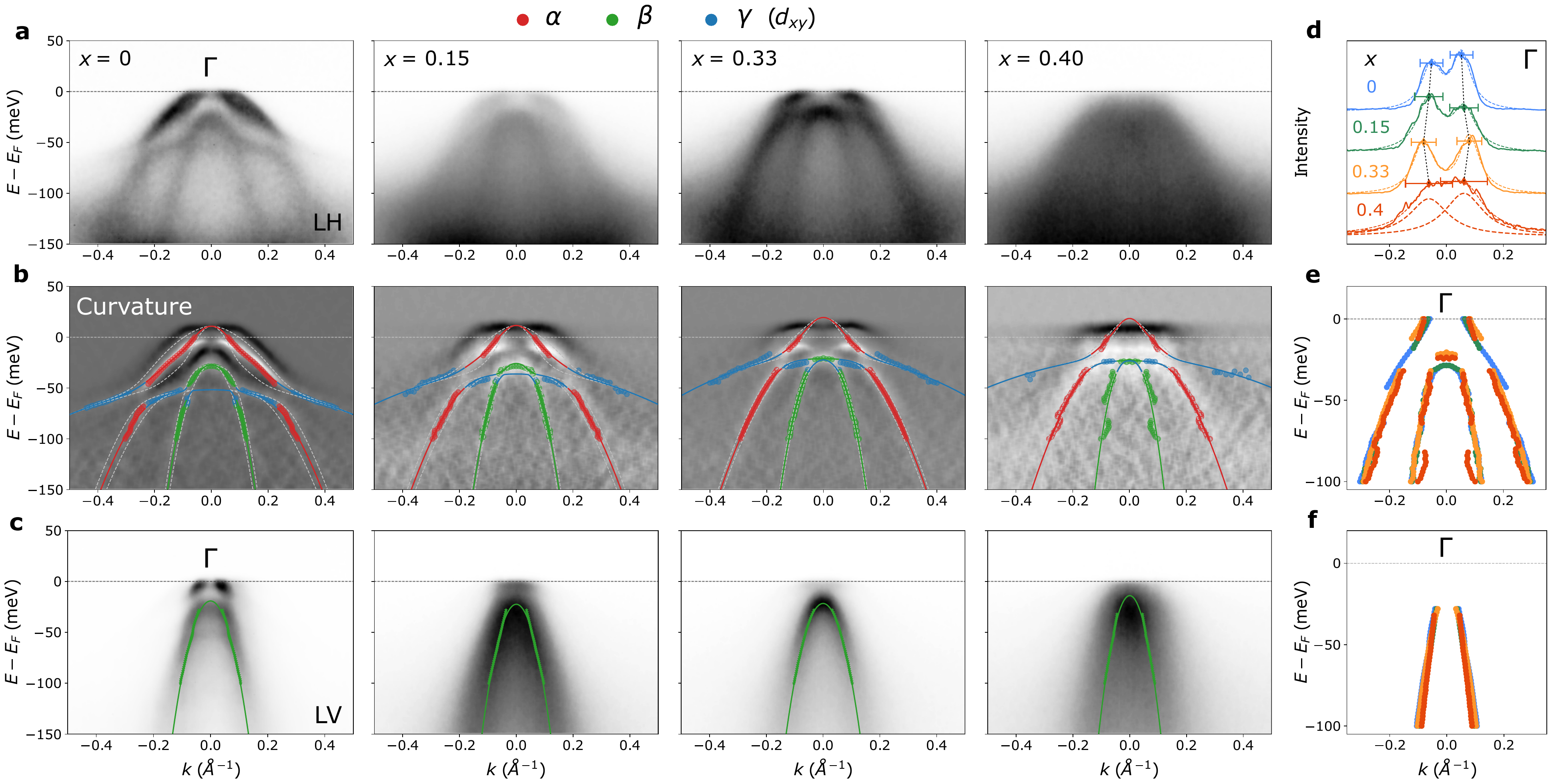}
    \caption{\textbf{Hole bands of nematic FeSe$_{1-x}$Te$_x$}.
    \textbf{(a)} Angle-resolved photoemission spectroscopy (ARPES) spectra measured at the $\Gamma$ high-symmetry point along the $M-\Gamma-M$ direction at 10~K using linear horizontal (LH) light polarisation.
    \textbf{(b)} The corresponding two-dimensional curvature of the ARPES data from \textbf{(a)} using the method in Ref. \cite{Zhang2011}.
    The solid points are the band positions extracted from Lorentzian fits to
    momentum distribution curves (MDCs) and extrema within constant momentum cuts of the raw data and the curvature. The solid lines are fits to the band positions using the low-energy model discussed in Supplementary Materials.
    \textbf{(c)} ARPES spectra for the same dispersions as in \textbf{(a)} but measured using LV light polarisation.
    The solid points are the band positions extracted from MDC analysis, and the solid curve is a parabolic fit to the data.
    The solid lines in \textbf{(b)} and \textbf{(c)} correspond to the outer hole band, $\alpha$ (red), in inner hole band, $\beta$ (green) and flatter $d_{xy}$ hole band, $\gamma$ (blue).
    \textbf{(d)} MDCs taken at the Fermi level are fitted to  Lorentzian curves   to determine the Fermi wavevector ($k_{\rm F}$) (the error is equal to half of the full-width half-maximum). The MDCs are vertically shifted for clarity.
      \textbf{(e)} The extracted hole dispersions of the $\alpha$ and $\beta$ bands  for different $x$ using LH polarisation. \textbf{(f)} The extracted $\beta$ bands using LV polarisation.
   Different colours  in \textbf{(d)}, \textbf{(e)} and \textbf{(f)} refer to the $x$ compositions as indicated by the key in \textbf{(d)}.
   }
    \label{SupFig_Gamma}
\end{figure}

To constrain the different parameters and to estimate of the SOC terms, this low-energy model was applied to the ARPES data in four separate ways.
Supplementary Figure~\ref{SupFig_Diff_Model} presents the values of $\varphi_{nem}$, $\lambda_1$ and $\lambda_2$ for the different cases, where the extracted $\lambda$ terms remain consistent between the cases.

As the Te content increases the linewidths broaden, resulting in a larger uncertainty in the evaluated band dispersions and the corresponding nematic splitting. Due to these effects, the model applied to FeSe$_{1-x}$Te$_x$ data becomes insensitive to the nematic order parameter, $\varphi_{nem}$ and it was fixed to be proportional to $T_{\rm s}$. Moreover, the value of $\lambda_2$ seemed relatively constant for each approach and is assumed to not drastically effect the key model parameters, subsequently in the fourth and final case it was fixed to an average value of 64~meV$\cdot$\AA.
For $\varphi \propto T_{\rm s}$, the value of $\lambda_1$ gradually increases with increasing the Te concentration. This is in broad agreement with the predictions of DFT
where the SOC splitting between the $\alpha$ and $\beta$ bands increases from 93 to 101~meV at the $\Gamma$
from FeSe towards FeTe.

Lastly, as the experimental data is twinned, the extracted band positions are an average of the two domains. Therefore, the employed model takes the average of equation \ref{Ham} when $k = k_x$ and $k_y$, such that both domains are combined.

\clearpage
\newpage
\subsection*{The effect  of the $p_z$ band}
\label{Appendix: pz}

The band inversion due to the crossing of the $p_z$ chalcogen band with the $\beta$ band is illustrated in Supplementary Figure~\ref{SupFig_DFT_ch}.
This effect is predicted to first occur at the $Z$ point when the chalcogen height is set to $h \sim 1.4$~\AA, however experimentally it was suggested that this inversion takes place for FeSe$_{0.5}$Te$_{0.5}$ \cite{Lohani2020, Wang2015, Chen2020}.
As the $p_z$ and $\beta$ band interact, they exchange their orbital character, which in turn affects the matrix elements and thus the intensity of the ARPES spectra.
Therefore, to quantify the presence of the $p_z$ band, we assess the spectral intensity of the $\beta$ band relative to the $\alpha$ band at both high-symmetry points points ($I(\beta,\alpha)_\Gamma$ and $I(\beta,\alpha)_Z$). To perform this analysis we integrate the intensity between two spectral regions which relate to the $\beta$ and $\alpha$ bands (see Supplementary Figure~\ref{SupFig_pz}b and c).
Lastly, by taking the ratio between the $Z$ and $\Gamma$ points, effects that are $k_z$ independent, such as nematic effects and the hybridisation with the $\gamma$ band, are cancelled out.

We find that the relative intensity of the $\beta$ band increases at the $Z$ point with respect to the $\Gamma$ point as the Te concentration increases beyond $x=0.33$, as shown in Supplementary Figure~\ref{SupFig_pz}d. The upturn observed for $x>0.33$ could suggest the influence of the $p_z$ band as it begins to hybridise with the $\beta$ band.
A similar analysis was performed in Ref.~\cite{Lohani2020,Li2023}, where the spectral intensity of the $\beta$ band was evaluated relative to a surface state. The $k_z$ dependence of this relative intensity was then used to establish the presence of the $p_z$ band.
Furthermore, the spectral intensity of the $\beta$ band, relative to the $\alpha$ band, is seen to vary with the Te concentration
by comparing the MDCs curves taken at -70~meV and -90~meV (Supplementary Figure~\ref{SupFig_pz}e and f respectively).

\begin{figure}[h]
    \centering
           \includegraphics[width=0.9\linewidth]{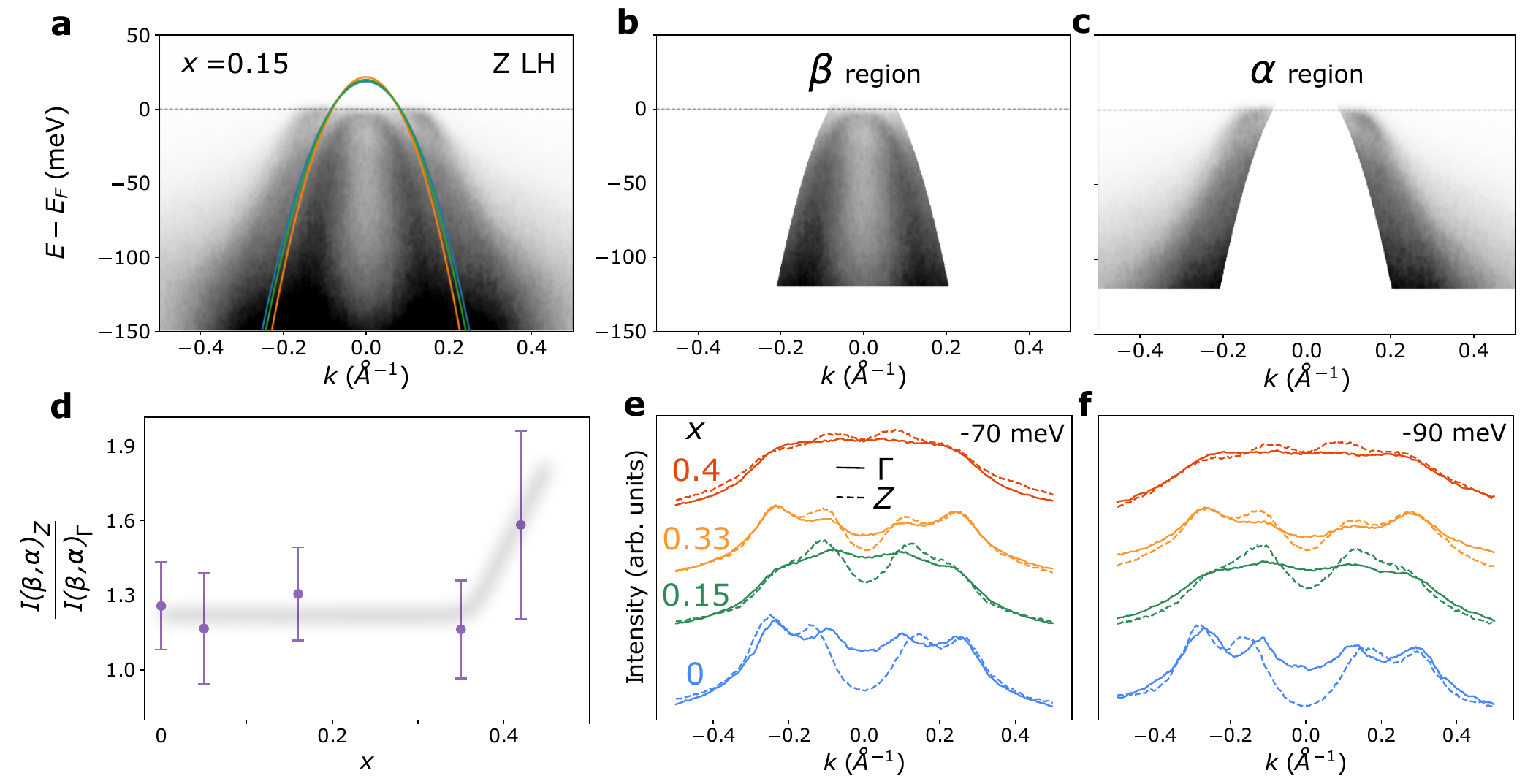}
    \caption{{\bf Quantifying the relative changes in intensity induced by the presence of the chalcogen $p_z$ band.}
    \textbf{(a)} The angle-resolved photoemission spectroscopy spectra for the hole pockets centered at the $Z$ point for $x=0.15$ measured in the  LH configuration.
    Parabolic curves are overlaid which act as a partition between the $\alpha$ and $\beta$ hole bands.
    \textbf{(b)} and \textbf{(c)} are the $\beta$ and $\alpha$ spectra formed by the bisecting parabola respectively,
    where the spectra is also cut off between -120meV and the top of the parabola. Each section is then integrated and the ratio of the $\beta$ band region and the $\alpha$ band region is extracted. At each composition this ratio $I(\beta,\alpha)_\Gamma$ and $I(\beta,\alpha)_\Gamma$ is determined.
     \textbf{(d)} The variation of $\frac{I(\beta,\alpha)_Z}{I(\beta,\alpha)_\Gamma}$ as a function of $x$.
    \textbf{(e)} and \textbf{(f)} Momentum distribution curves (MDCs) taken at -70meV and -90meV respectively for all compositions (as defined by the colours),
    where the MDCs are vertically offset for clarity. The solid curves correspond to the $\Gamma$ point and the dashed curves to the $Z$ point.
    }
    \label{SupFig_pz}
\end{figure}

\clearpage
\newpage

\subsection*{The evolution of the bands with $d_{z^2}$ orbital character }
\label{Appendix: dz2}

\begin{figure}[h]
    \centering
    \includegraphics[width=0.9\linewidth]{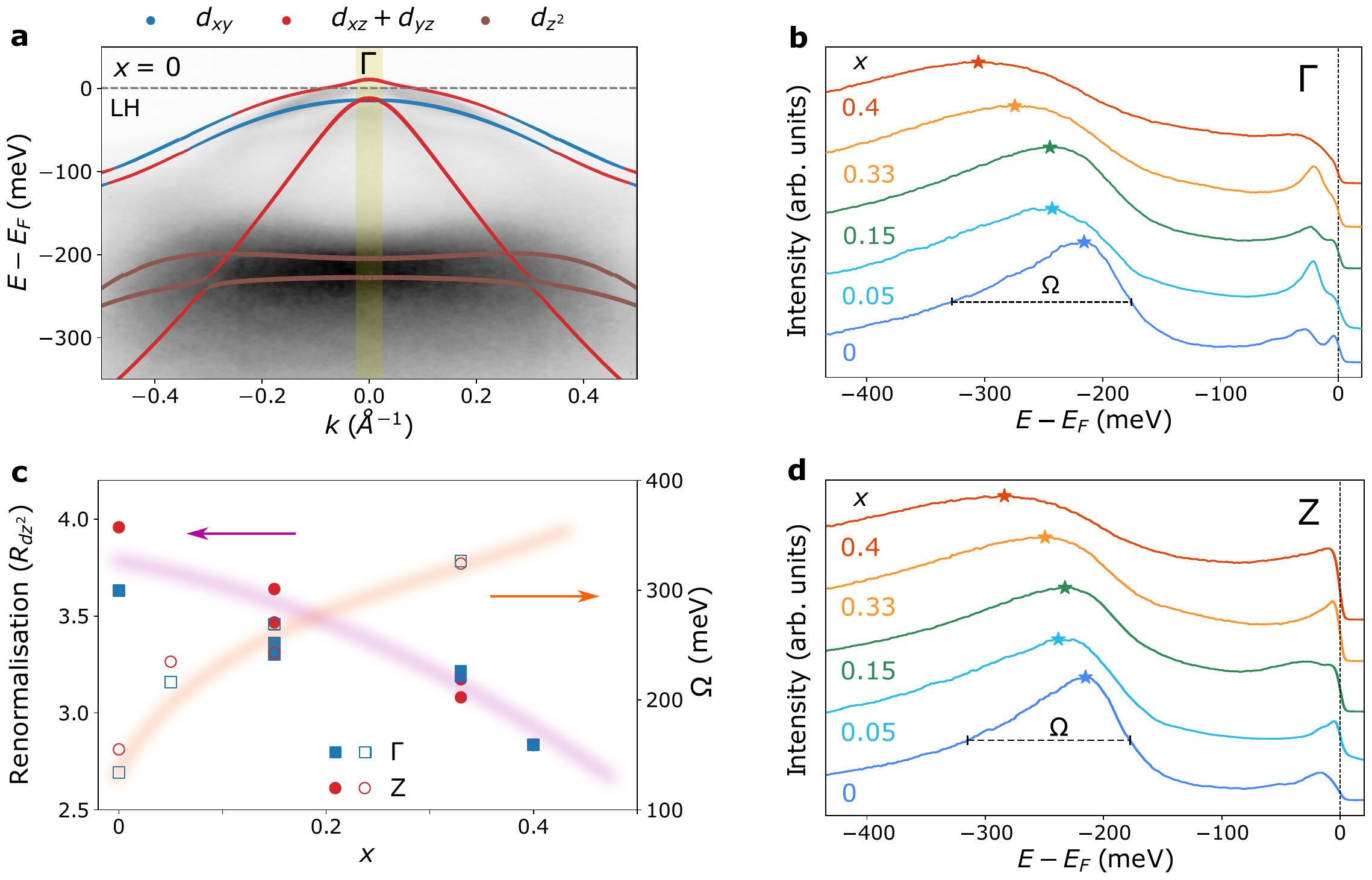}
          \caption{\textbf{The evolution of the $d_{z^2}$ hole bands.}
    \textbf{(a)} Angles-resolved photoemission spectroscopy data along the $M-\Gamma-M$ dispersion for FeSe at 10~K using LH polarisation, taken from Ref.~\cite{Watson2015a}. The solid curves correspond to the density functional theory calculations which are renormalised by a factor 3.9 and shifted by 20meV. The colours of the bands correspond the dominant orbital character of each band.
   \textbf{(b)} The energy distribution curves (EDCs) for the different FeSe$_{1-x}$Te$_x$ samples at the $\Gamma$ point, which correspond to the intensity profile along the yellow strip in \textbf{(a)}
   with an integration window of $0.05 $\AA$^{-1}$. The EDC curves are offset vertically for clarity. For $x=0$ the full-width half-maximum is defined as $\Omega$.
   \textbf{(c)} Renormalisation factor,  $R_{dz^2}$,  estimated at the $\Gamma$ (solid red circle) and $Z$ (solid blue squares) high-symmetry points plotted on left y-axis. On the right y-axis the broadening, $\Omega$ (open symbols), is plotted against concentration, $x$. The solid lines are guides to the eye.
   \textbf{(d)} The same plot as \textbf{(b)} but for the $Z$ high symmetry point.
   }
    \label{SupFig_dz2_Part1}
\end{figure}

Here we focus on the behaviour of the $d_{z^2}$ bands located at the Brillouin zone centre at higher binding energy.
Fig~\ref{SupFig_dz2_Part1}a shows a broad feature at $\sim$220~meV in the ARPES spectra of FeSe centred around the $\Gamma$ point.
While only one spectral feature can be resolved in the experiments, DFT calculations predicts two different dispersion corresponding to the bands with $d_{z^2}$ orbital character.
We estimate that a renormalisation factor of 3.9 brings the calculations in agreement with the experimental feature of FeSe, as shown in Supplementary Figure~\ref{SupFig_dz2_Part1}a.
Supplementary Figure~\ref{SupFig_dz2_Part1}b and d shows that the spectral feature of the $d_{z^2}$ bands (as determined from EDCs centred at $\Gamma$ and the $Z$ point)
shifts towards higher binding energy and broadens with the Te substitution (see Supplementary Figure~\ref{SupFig_dz2_Part1}c),
in agreement with a previous report \cite{Liu2015}.
Although this broadening, $\Omega$, is likely to occur due to the splitting of the two bands, as predicted by the DFT calculations
(see Supplementary Figure~\ref{SupFig_dz2_Part2}b), other mechanisms are expected to also contribute.
Firstly, as bands move to higher binding energy there is a natural broadening of the spectral function
\cite{Damascelli2004}, and secondly broadening caused by the impurity scattering and disorder is likely to increase as
a function of Te concentration.
 To estimate the renormalisation factor of the $d_{z^2}$ bands ($R_{dz^2}$),
we assess the ratio between the energy separation between the top of the $\alpha$ hole band and the average position of the $d_{z^2}$ bands from calculations and experiments.
DFT predicts that the average position of the two $d_{z^2}$ bands should also shift to higher binding energies as Te concentration is increased, yet we find that $R_{dz^2}$ still decreases as a function of Te composition, as shown in Supplementary Figure~\ref{SupFig_dz2_Part1}c.

\begin{figure}[h]
    \centering
    \includegraphics[width=0.7\linewidth]{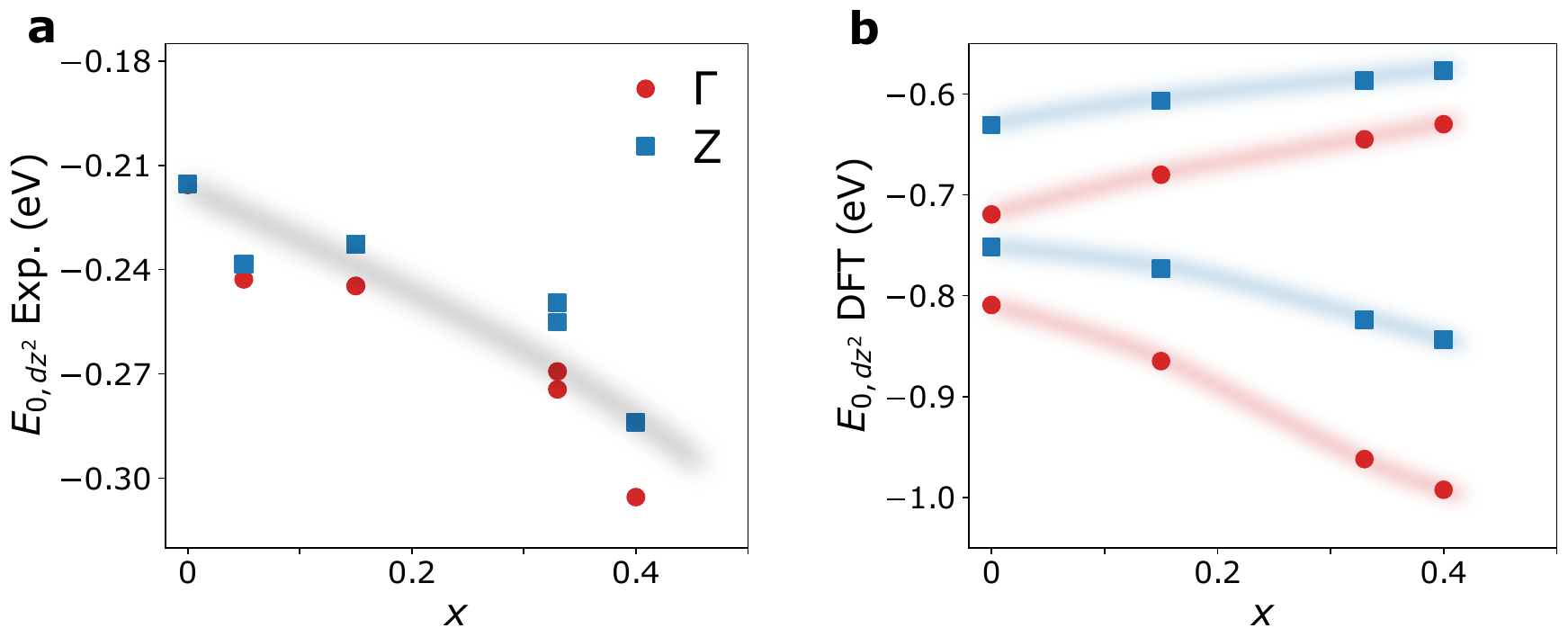}
       \caption{\textbf{Shifts of the $d_{z^2}$ bands.}
    \textbf{(a)} The changes in $d_{z^2}$ spectral feature  as a function of composition $x$, determined from the EDCs.
    \textbf{(b)} The  changes in the position of the two bands with $d_{z^2}$ orbital character, as determined from the DFT series in Supplementary Figure~\ref{SupFig_DFT}.
    For all plots red solid circles correspond to data taken at $\Gamma$ and blue solid squares correspond to $Z$. The solid lines are guides to the eye. }
    \label{SupFig_dz2_Part2}
\end{figure}

\newpage
\clearpage

\subsection*{ Quantifying the nematic order affecting the electron pockets}
\label{Appendix: Electron}

\begin{figure}[h]
    \centering
        \includegraphics[width=\linewidth]{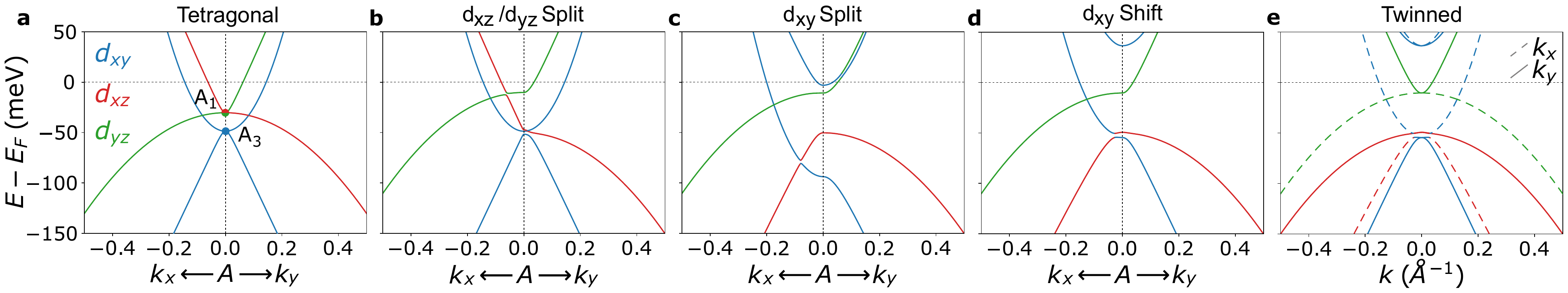}
        \caption{\textbf{The effect of the nematic order on the electron pockets}.
        \textbf{(a)} Band dispersion at the Brillouin zone corner in the tetragonal phase where there are two degenerate points, one formed by the $d_{xz/yz}$ bands (A1) and the other from the $d_{xy}$ bands (A3).
        \textbf{(b)} Band dispersion where the degeneracy of the $d_{xz/yz}$ bands is lifted.
        \textbf{(c)} Band dispersion where both the degeneracy of the $d_{xz/yz}$ bands and the $d_{xy}$ bands are lifted \cite{Fernandes2014a}.
        \textbf{(d)} Band dispersion where both degeneracies are lifted and an additional global shift of the $d_{xy}$ bands is applied based on a proposal from Ref.~\cite{Rhodes2021}.
        \textbf{(e)} Twinned spectrum of dispersion given in \textbf{(d)}.
        For all plots the colours of the bands refer to the dominant orbital character ($d_{xy}$ - blue, $d_{xz}$ - red and $d_{yz}$ - green).
        }
     \label{SupFig_Electron_nematic}
\end{figure}

Supplementary Figure~\ref{SupFig_Electron_nematic}a shows the band structure of tetragonal FeSe where there are two degenerate points, one associated with the $d_{xz}$ and $d_{yz}$ bands (A1) and one with $d_{xy}$ bands (A3). As the nematic order manifests the degeneracy of the $d_{xz/yz}$ bands should be lifted, conforming to broken crystal symmetry (see Supplementary Figure~\ref{SupFig_Electron_nematic}b).
In contrast, the $d_{xy}$ orbital does not conform to the broken symmetry channel, yet an anisotropic hopping of this orbital still lifts the degeneracy (see Supplementary Figure~\ref{SupFig_Electron_nematic}c) \cite{Fernandes2014a, Rhodes2021, Yi2019}.
To probe the nematic order, previous studies have measured the separation between spectral features along an EDC taken at the Brillouin corner, $\Delta_{M/A}$, where this separation was
assigned to different types of orbital effects \cite{Yi2019, Fernandes2014a, Christensen2020,Coldea2021}.

\begin{figure*}[htbp]
    \centering
        \includegraphics[width=\linewidth]{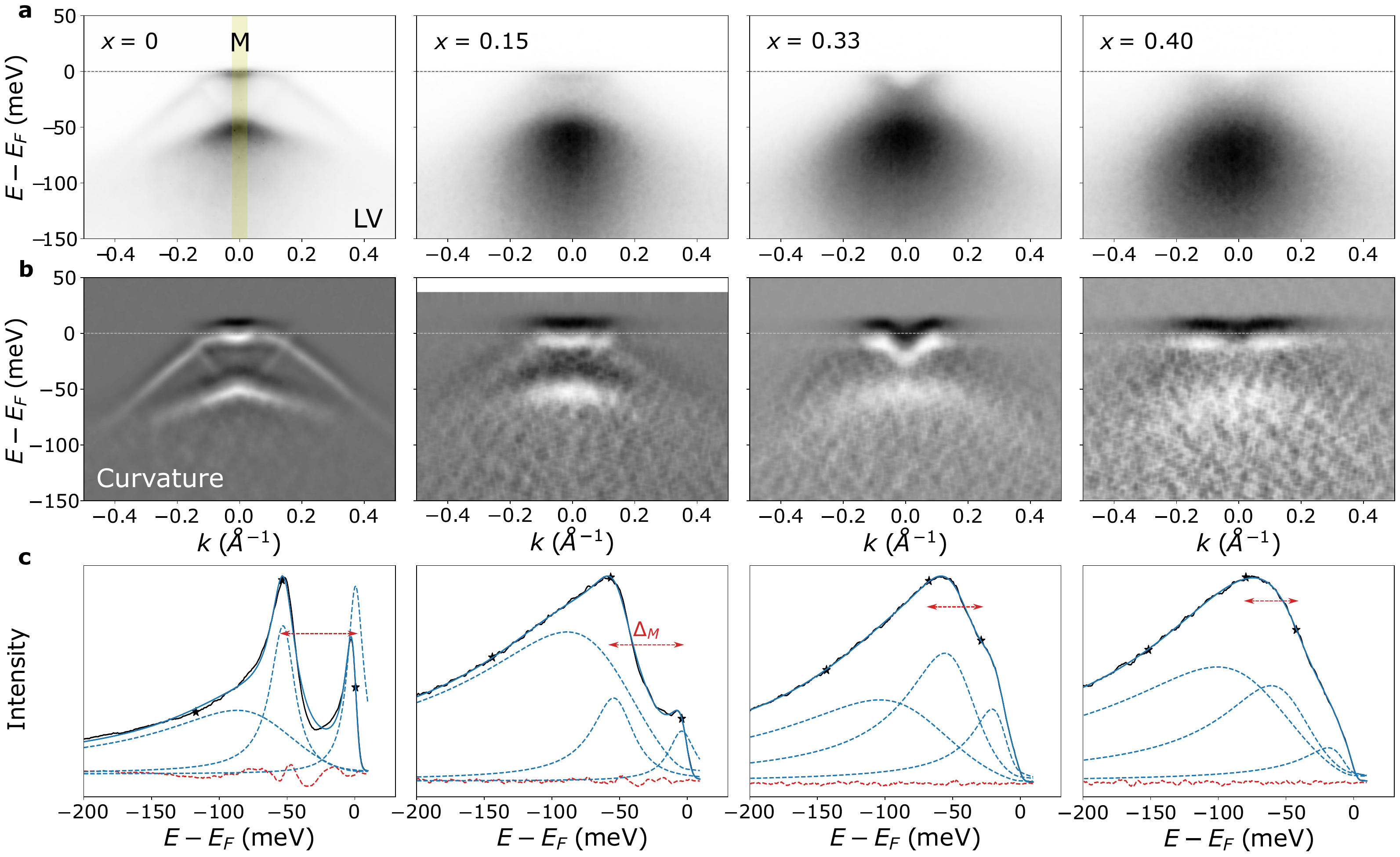}
       \caption{\textbf{Electron Pocket of FeSe$_{1-x}$Te$_x$.}
    \textbf{(a)} Angle-resolved photoemission spectroscopy spectra taken at 10~K along the $\Gamma-M-\Gamma$ direction using linear vertical light polarisation for the different compositions $x$.
        \textbf{(b)} The two-dimensional curvature of the ARPES spectra in \textbf{(a)}.
    \textbf{(c)} The estimation of $\Delta_M$ by fitting the energy distribution curves (solid black curve) (integrated within the window indicated by the yellow strip for $x=0$ in \textbf{(a)}).
     The blue solid curve is the fit, the dashed blue curves are the individual Lorentzian curves and dashed red plot is the residuals. Star data points correspond to the position of the Lorentzian centre.
        }
    \label{SupFig_M_point}
\end{figure*}

In order to quantify the separation in the spectral features, $\Delta_{M/A}$,
we fit the EDC spectra to Lorentzian curves with a broadening that is linearly dependent on the binding energy  using the following expression:
\begin{equation}
	I_{EDC} = \Sigma_i A_i \frac{\Gamma_i/2}{(x-x_i)^2 + (\Gamma_i/2)^2}
\end{equation}
$$ \Gamma_i = \Gamma_0 + a_i E $$
where $i$ is the Lorentzian index such that $A_i$, $x_i$ and $\Gamma_i$ are the amplitude, offset and broadening of the $i^{\text{th}}$ Lorentzian curve respectively.
The broadening contains an energy independent term, $\Gamma_0$, which is fixed for all Lorentzians,
as this would represent a scattering term due to impurities, and an energy dependent term with $a_i$ as the coefficient.
This linear dependence
 is a hallmark of a marginal Fermi-liquid, found in other Fe-based superconductors \cite{Fink2021}.
Supplementary Figure~\ref{SupFig_M_point}c
shows the fitting procedure to assess $\Delta_{M}$, where for all compositions, three different peaks are used and $\Delta_{M}$ is the separation between the two top most peaks.
One should also note that the maximum of a Lorentzian with an energy dependent broadening will not be equal to the centre of the Lorentzian, as clearly shown in Supplementary Figure~\ref{SupFig_M_point}c.

\end{document}